# Lattice dynamics study in PbWO$_4$ under high pressure

F.J. Manjón,[1,2,*] D. Errandonea,[1] N. Garro,[1] J. Pellicer-Porres,[1] J. López-Solano,[3] P. Rodríguez-Hernández,[3] S. Radescu,[3] A. Mujica,[3] and A. Muñoz[3]

[1] Dpto. de Física Aplicada i Institut de Ciencia de Materials de la Universitat de València, 46100 Burjassot (València), Spain
[2] Dpto. de Física Aplicada, Universitat Politècnica de València, 46022 València, Spain
[3] Dpto. de Física Fundamental II, Universidad de La Laguna, La Laguna 38205, Tenerife, Spain

**Abstract.** Room-temperature Raman scattering has been measured in lead tungstate up to 17 GPa. We report the pressure dependence of all the Raman modes of the tetragonal scheelite phase (PbWO$_4$-I, space group I4$_1$/$a$), which is stable at ambient conditions. Upon compression the Raman spectrum undergoes significant changes around 6.2 GPa due to the onset of a partial structural phase transition to the monoclinic PbWO$_4$-III phase (space group P2$_1$/$n$). Further changes in the spectrum occur at 7.9 GPa, related to a scheelite-to-fergusonite transition. This transition is observed due to the sluggishness and kinetic hindrance of the I→III transition. Consequently, we found the coexistence of the scheelite, PbWO$_4$-III, and fergusonite phases from 7.9 to 9 GPa, and of the last two phases up to 14.6 GPa. Further to the experiments, we have performed *ab initio* lattice dynamics calculations which have greatly helped us in assigning the Raman modes of the three phases and discussing their pressure dependence.

---

[*] Corresponding author with present address at UPV. E-mail address: fjmanjon@fis.upv.es
Tel.: + 34 96 387 52 87, Fax: + 34 96 387 71 89

**I. Introduction**

Lead tungstate (PbWO$_4$) has been proposed as an excellent material for the implementation of Raman lasers due to the strong scattering cross section of the highest A$_{1g}$ mode of the scheelite structure [1]. It is also one of the candidate materials chosen for the cryogenic phonon-scintillation detectors at the high-energy electromagnetic calorimeter of the Large Hadron Collider at CERN [2,3]. For both applications a detailed knowledge of its lattice dynamics and structural properties is highly desirable.

PbWO$_4$ is a compound that crystallizes in the tetragonal scheelite-type structure (PbWO$_4$-I or stolzite, space group [SG]: I4$_1$/a, No. 88, Z = 4) at ambient conditions [4]. However, other two metastable polymorphs have also been observed: raspite-type (PbWO$_4$-II, SG: P2$_1$/a, No. 14, Z = 4) [5,6], and PbWO$_4$-III (SG: P2$_1$/n, No. 14, Z = 8) [7]. The raspite phase coexists with the scheelite phase in some natural samples. On the other hand, the PbWO$_4$-III phase can only be obtained after a high-pressure high-temperature treatment and several authors have proposed this phase as a candidate high-pressure phase at room temperature (RT) in PbWO$_4$ [8,9].

As already commented in our recent Raman work in BaWO$_4$ under pressure [10], the behavior of scheelite-type tungstates under pressure has been studied since the 1970s. Recently a systematic study of their structure under high pressure has been performed in alkaline-earth tungstates and PbWO$_4$ by means of angle dispersive x-ray diffraction (ADXRD) and x-ray absorption near-edge structure (XANES) measurements in powder samples, complemented with *ab initio* total-energy calculations [11,12]. It has been found that scheelite PbWO$_4$ transforms under pressure to the monoclinic M-fergusonite structure (hereafter called fergusonite, SG: I2/a, No. 15, Z = 4) at 9.0 GPa [12]. This transition pressure is in good agreement with the value estimated for PbWO$_4$ from the correlation of the transition pressures with the packing ratio of anionic BX$_4$ units around A cations in scheelite-related ABX$_4$ compounds [13]. In **Ref. 12** it was also found that PbWO$_4$ undergoes a second phase transition to the monoclinic PbWO$_4$-III phase around 15 GPa. However, there is still some controversy concerning the high pressure phases of PbWO$_4$. *Ab initio* total-energy calculations indicated that PbWO$_4$-III is the energetically most favored structure beyond 5.3 GPa and should be the only phase observed at least up to 20 GPa because the ferguson ite structure is more stable than scheelite structure only beyond 8 GPa [12]. The controversy is even more evident if we consider that an early Raman study [9] and a recent ADXRD study [14] in PbWO$_4$ showed indications of a phase transition at 4.5 GPa and 5 GPa, respectively, but there are some puzzling questions regarding these works. In **Ref. 14**, the ADXRD patterns have been assigned to stolzite up to 10 GPa, despite the observation of changes at 5 GPa. Furthermore, the observed new weak reflections cannot be accounted either by the ferguson ite or the PbWO$_4$-III structures. On the other hand, in the Raman study [9], it was suggested that the high-pressure phase of PbWO$_4$ could be different than that of BaWO$_4$ because of the different Raman mode frequencies observed in the high-pressure phase of both materials. However, no Raman spectrum of the high pressure phase of PbWO$_4$ was provided to compare it with that of BaWO$_4$. On top of that, some Raman modes of the scheelite phase were not found in that work. Therefore, the lack of high-pressure Raman spectra in PbWO$_4$, the absence of some first-order modes of the scheelite structure, and the apparent discrepancies among earlier experimental and theoretical works justify our present study.

As part of our project to study the stability of scheelite-structured tungstates and to give a comprehensive description of their complex high-pressure phase diagrams, we report in this work a RT Raman study of PbWO$_4$ up to 17 GPa. In addition, we present *ab initio* lattice dynamics calculations that have assisted us in the assignment and discussion of the behavior of the zone-center phonons in the different structural phases. This work complements a recent high-pressure Raman work in BaWO$_4$ [10], in which it was shown that in BaWO$_4$ the onset of a partial scheelite-to-BaWO$_4$-II phase transition occurs at 6.9 GPa; i.e., at lower pressure than the scheelite-to-ferguson ite transition, which was observed at 7.5 GPa. These results are in good agreement with previous *ab initio* total-energy calculations and ADXRD and XANES measurements **[12]**. Furthermore, previous high-pressure Raman spectra in BaWO$_4$ **[8]** were interpreted on the basis of the results of **Ref. 10**. Similarly, we will show in this work that PbWO$_4$ suffers the same phase transitions than BaWO$_4$ and that the frequencies of the Raman modes in the high-pressure phases of PbWO$_4$ previously reported **[9]** can be completely understood on the light of the present work. Our results allow us to develop a picture of the structural behavior of PbWO$_4$ that solves apparent discrepancies among earlier experiments and theory.

**II. Experimental details**

The PbWO$_4$ samples used in this study were obtained from scheelite-type bulk single crystals which were grown with the Czochralski method starting from raw powders having 5N purity **[15,16]**. Small platelets (100μm x 100μm x 30μm) were cleaved from these crystals along the {101} natural cleavage plane **[17]** and inserted in a diamond-anvil cell (DAC). Silicone oil was used as pressure-transmitting medium **[18]** and the pressure was determined by calibration with the ruby photoluminescence **[19]**. Raman measurements at RT were performed in backscattering geometry using the 488 Å line of an Ar$^+$-ion laser with a power of less than 100 mW before the DAC. Dispersed light was analyzed with a Jobin-Yvon T64000 triple spectrometer equipped with a confocal microscope in combination with a liquid nitrogen (LN)-cooled multi-channel CCD detector. Spectral resolution was better than 1 cm$^{-1}$ and Ar and He plasma lines were used to calibrate the Raman and photoluminescence spectra.

**III. *Ab initio* lattice dynamics calculations**

Along with the experimental data of our Raman study we will also present results of a theoretical *ab initio* calculation of the phonon modes of the scheelite, fergusonite, and PbWO$_4$-III phases at the zone center (Γ point). All the calculations were done within the framework of the density functional theory (DFT) using the Vienna *ab initio* simulation package (VASP) **[20]**. The exchange and correlation energy was taken in the generalized gradient approximation (GGA) according to Perdew-Burke-Ernzerhof (GGA-PBE). The projector-augmented wave (PAW) scheme **[21]** was adopted and the semicore 5d electrons of Pb were dealt with explicitly in the calculations. The set of plane waves used extended up to a kinetic energy cutoff of 625 eV. We perform highly converged calculations in order to obtain the dynamical matrix as described in **Ref. 10**.

**IV. Results and discussion**

**A. Low-pressure phase: Scheelite structure**

PbWO$_4$ crystallizes at ambient conditions in the centrosymmetric scheelite structure that has space group I4$_1$/a (C$^6_{4h}$), with four formula units per body centered unit cell. The Pb and W atoms occupy S$_4$ sites whereas the sixteen oxygen atoms are on general C$_1$ sites. Group theoretical considerations **[22-24]** lead us to expect 13 Raman-active modes at the Γ point **[10]**:

$$\Gamma = \nu_1(A_g) + \nu_2(A_g) + \nu_2(B_g) + \nu_3(B_g) + \nu_3(E_g) + \nu_4(B_g) + \nu_4(E_g)$$
$$+ R(A_g) + R(E_g) + 2T(B_g) + 2T(E_g) \qquad (1)$$

The translational modes (T) and the rotational modes (R) are considered to be the external modes of the WO$_4$ tetrahedra and are the lowest in frequency. The rest ($\nu_1$ to $\nu_4$) are considered to be the internal modes of the WO$_4$ tetrahedra and higher in frequency. The $A_g$ and $B_g$ modes are single, while $E_g$ modes are doubly degenerated.

To the best of our knowledge twelve of the thirteen modes of the scheelite-type phase of PbWO$_4$ are known **[25,26]**, the $\nu_4(E_g)$ internal mode being the only unknown one. This mode has been observed in CaWO$_4$, SrWO$_4$, and BaWO$_4$ as a high-frequency shoulder of the $\nu_4(B_g)$ internal mode **[10,27,28]**. Only the pressure dependence of ten of the thirteen modes is known **[9]**. **Fig. 1** shows the RT Raman spectra of stolzite at different pressures up to 8 GPa. The Raman spectra should correspond to a mixture of polarizations perpendicular and parallel to the c-axis because of our sample orientation. In order to assign the different Raman modes of stolzite we have followed the notation of Liegeois-Duyckaerts and Tarte **[29]**. Marks at the bottom of **Fig. 1** indicate the *ab initio* calculated frequencies of the Raman modes in scheelite-PbWO$_4$ at 1 atm. It can be seen that our experimental and theoretical Raman frequencies at 1 atm compare reasonably well. The Raman spectrum of stolzite is dominated by the $\nu_1(A_g)$ mode near 906 cm$^{-1}$ at 1 atm; i.e., the mode used in Raman lasers. In addition, one can distinguish clearly at least other ten modes in the experimental Raman spectrum of the scheelite phase. Only two modes (one $\nu_2$ and one $\nu_4$ mode) cannot be clearly observed because they probably overlap with other modes, as can be inferred from the proximity of the calculated $\nu_2$ and $\nu_4$ frequencies (see bottom of **Fig. 1**). A closer inspection of the modes located near 328 and 357 cm$^{-1}$ at 1 atm allows us to conclude that they are indeed double modes. Inset of **Fig. 1** shows a detailed Raman spectrum of the region near 357 cm$^{-1}$ at 1 atm measured outside the DAC. We have found a mode at 363 cm$^{-1}$ at 1 atm as a high-frequency shoulder of the $\nu_4(B_g)$ mode near 357 cm$^{-1}$. In a similar fashion, we have found a mode at 323 cm$^{-1}$ at 1 atm as a low-frequency shoulder of the $\nu_2(B_g)$ mode located at 328 cm$^{-1}$ at 1 atm. **Fig. 2** shows a detail of the Raman spectra near the $\nu_2(B_g)$ mode at several pressures. The spectra at different pressures have been shifted in frequency in order to bring the $\nu_2(B_g)$ mode into coincidence, so that the relative shift of the low-frequency shoulder with respect to the $\nu_2(B_g)$ mode as a function of pressure can be observed. The low-frequency tail of the anisotropic $\nu_2(B_g)$ mode linewidth becomes more pronounced with increasing pressure. This feature can be attributed to the presence of the $\nu_2(A_g)$ mode at the low frequency side of the $\nu_2(B_g)$ mode. In summary, we have tentatively assigned the modes located at 323 and 367 cm$^{-1}$ to the $\nu_2(A_g)$ mode and the unreported $\nu_4(E_g)$ mode in scheelite-type PbWO$_4$, respectively. The assignment of the $\nu_2(A_g)$ mode is based on the ordering of the two modes and on the slightly smaller pressure coefficient of the $\nu_2(A_g)$ mode respect to the $\nu_2(B_g)$ mode, as indicated by our lattice dynamics calculations, and already commented in the case of scheelite-BaWO$_4$ **[10]**. The assignment of the $\nu_4(E_g)$ mode is based on its location, on its similar pressure coefficient to that of the $\nu_4(B_g)$ mode, according to our calculations, and on the similar location of this mode in CaWO$_4$, SrWO$_4$, and BaWO$_4$ **[10,27,28]**. Further support for the assignment of the previously unobserved $\nu_2$ and $\nu_4$ modes, comes from the knowledge that the intensity of the $\nu_2$ modes must be higher than those of $\nu_4$ modes **[30]**, and that the lack of observation of several Raman $\nu_2$ and/or $\nu_4$ modes in tetrahedral ABO$_4$-type compounds is due to the fact that the $\nu_2$ and/or $\nu_4$ frequencies almost coincide **[31]**, as indeed found in our calculations.

**Figure 3** shows the pressure dependence of the Raman mode frequencies of stolzite (solid circles) up to 8 GPa. **Table I** summarizes the frequencies ($\omega$) of all the scheelite Raman modes, their pressure coefficients (d$\omega$/dP), and Grüneisen parameters ($\gamma = B_0/\omega \cdot d\omega/dP$, with $B_0$= 66 GPa being the stolzite bulk modulus **[12]**). In **Table I**, we also compare the experimental results for stolzite with those obtained from our calculations. For completeness, **Table II** summarizes the calculated frequencies and pressure coefficients of the infrared (IR) modes of stolzite that compare reasonably well to the experimental frequencies obtained from the literature **[25,26,32]**.

Our measured Raman frequencies, pressure coefficients, and Grüneisen parameters in scheelite-PbWO$_4$ agree with those reported by Jayaraman *et al.* **[9]**. The only significant deviation corresponds to pressure coefficients measured for the lowest T(E$_g$) mode at 65 cm$^{-1}$ and the T(B$_g$) mode at 77 cm$^{-1}$. We have measured a much smaller pressure coefficient (1.8 cm$^{-1}$/GPa) for the T(E$_g$) mode than that measured previously (3.4 cm$^{-1}$/GPa) **[9]**. Consequently, our pressure coefficient gives a Grüneisen parameter γ of 1.9 that is much smaller than the 3.2 obtained by Jayaraman *et al*. **[9]**. Note that there is a mistake in Table I of **Ref. 9** and the reported Grüneisen parameters of PbWO$_4$ and PbMoO$_4$ are underestimated by an order of magnitude. Our pressure coefficient for the lowest T(E$_g$) mode is similar to those in other scheelite tungstates (between 1 and 1.7 cm$^{-1}$/GPa) and our γ is in agreement with those found for this mode in other scheelite tungstates (between 1 and 1.4) **[10]**. As regards to our γ for the T(B$_g$) mode at 77 cm$^{-1}$ (2.8), it seems to be rather high as compared to the same mode in other scheelites **[8,9,10]**, but the pressure coefficient (3.26 cm$^{-1}$/GPa) is remarkably similar to that recently measured for the highest T(B$_g$) phonon (3.4 cm$^{-1}$/GPa) in SrWO$_4$ **[27],** and even smaller than the same mode in CaWO$_4$ (4.7 cm$^{-1}$/GPa) **[28]**. It must be noted that this mode was not observed in the oldest Raman studies on scheelite tungstates under pressure **[33,34]**. The small γ for this mode in stolzite is due to the strong decrease of its frequency at 1 atm in the tungstate series (Ca, Sr, Ba, Pb) **[29]**. In fact, the Grüneisen parameters for the three lowest frequency modes in stolzite are much larger than those of the same modes in alkaline-earth scheelites because of the smaller frequencies of those modes in PbWO$_4$. In fact, the frequencies of all the external T modes in scheelite tungstates are inversely proportional to the square root of the cation mass due to the negligible contribution of the $WO_4$ tetrahedron to the frequency of these modes, and consequently decrease smoothly in the Sr, Ba, Pb series. However, this not the case for the highest $T(B_g)$ phonon which suffers an exceptional decrease in frequency from 133 cm$^{-1}$ in $BaWO_4$ to 77 cm$^{-1}$ in $PbWO_4$ **[29]**.

One common assumption in $ABO_4$–type scheelites regarding the pressure coefficients of their Raman modes is that the relatively stable tetrahedral $BO_4$ units are not as affected by pressure as dodecahedral $AO_8$ units, and consequently the pressure coefficients of the $WO_4$ tetrahedra internal modes should be smaller than those of the external modes. In this sense, we must note that $PbWO_4$ follows this trend more closely than alkaline-earth tungstates (see **Table I** in this work and in **Ref. 10**). It can be also observed that the pressure coefficients of all internal modes in $PbWO_4$ are smaller than those in the alkaline-earth tungstates. We believe that this result could be due to the smaller ionicity of the $PbWO_4$ with respect to the alkaline-earth tungstates, as discussed below.

It is known that the frequency of the stretching modes in $WO_4$ depends on the square root of the bonding force constant k, which increases with the intensity of the W-O interaction and decreases with the W-O bond distance. In general, a pressure increase should not alter very much the intensity of the W-O interaction, but reduces the bond distance resulting in an increase of the force constant, and consequently of the frequency. However, the W-O bond compressibility usually decreases with increasing the compound ionicity because the W-O bond compressibility decreases with increasing the charge transfer from the $A^{2+}$ cation to the $WO_4^{2-}$ anion. The consideration of the above statements and the observation of rather different pressure coefficients for the stretching $v_1(A_g)$ mode in the four scheelite $AWO_4$ tungstates (A = Ca, Sr, Ba, Pb) suggest a different behavior of the W-O interaction or of the W-O bond distance under pressure in the four tungstates, in particular between the most ionic (BaWO$_4$) and the least ionic compound (PbWO$_4$). Since BaWO$_4$ and PbWO$_4$ have similar W-O bond compressibilities [12,13], the only explanation for the factor 3 between the relative pressure coefficients of the stretching $\nu_1(A_g)$ mode in these two compounds is that it must be a change in the intensity of the W-O interaction in these tungstates with increasing pressure.

In order to complete the understanding of the pressure dependence of the Raman modes in stolzite in comparison with scheelite alkaline-earth tungstates, let us analyze further the high-frequency stretching modes of this phase. All the asymmetric stretching $\nu_3$ modes have similar frequencies and pressure coefficients in CaWO$_4$, SrWO$_4$ and BaWO$_4$ [10]. In these compounds, the frequency (pressure coefficient) is near 797 cm$^{-1}$ (3 cm$^{-1}$/GPa) and near 835 cm$^{-1}$ (2 cm$^{-1}$/GPa) for the $\nu_3(E_g)$ and the $\nu_3(B_g)$ modes, respectively. In PbWO$_4$, these frequencies are 5 to 8% smaller and their pressure coefficients are 25 to 50% smaller. On the other hand, the frequency and pressure coefficient of the symmetric stretching $\nu_1(A_g)$ mode increases from 911 to 926 cm$^{-1}$ and from 1.5 to 2.7 cm$^{-1}$/GPa, respectively when going from CaWO$_4$ to BaWO$_4$ [10]. This evolution suggests a dependence of the $\nu_1(A_g)$ mode on A cation parameters, despite this mode is an internal mode of the WO$_4$ tetrahedra and should be basically independent of the A cation, as already mentioned. The dependence of the frequency of the $\nu_1(A_g)$ mode on the A cation was confirmed by Dean *et al.* from Raman and IR measurements in aqueous solutions [35]. The comparison of the frequencies and pressure coefficients of the $\nu_1(A_g)$ mode in alkaline-earth tungstates with those in PbWO$_4$ suggests that the scaling of this mode does not depend on the A cation mass or ionic radius, but on the ionicity of the compound. The ionicity of the AWO$_4$ compound depends on the electronegativity of the A$^{2+}$ cation with respect to the WO$_4^{2-}$ anion.

Ordering the $A^{2+}$ cations in increasing electronegativity ($Ba^{2+}$, $Sr^{2+}$, $Ca^{2+}$, $Pb^{2+}$) correlates with the decrease of the frequency and pressure coefficient of the stretching $\nu_1(A_g)$ mode, thus indicating that the smaller ionicity of $PbWO_4$, as compared to the alkaline-earth tungstates, leads to smaller frequencies and pressure coefficients of the internal stretching vibrations of the $WO_4$ molecule in $PbWO_4$ than in alkaline-earth tungstates.

Additional support for the dependence of the stretching frequencies on the compound ionicity is obtained from the frequencies measured in several $BO_4$ molecules: 1) the $\nu_1$ and $\nu_3$ frequencies increase with the cation valence in the $[WO_4]^{4-}$ to $[WO_4]^{2-}$ series; 2) the $\nu_1$ and $\nu_3$ frequencies increase with the cation mass in the $[CrO_4]^{2-}$, $[MoO_4]^{2-}$, $[WO_4]^{2-}$ series; and 3) the $\nu_1$ and $\nu_3$ frequencies increase with the cation valence in the $[WO_4]^{2-}$, $[ReO_4]^-$, $[OsO_4]$ series, with W, Re, and Os belonging to the same row in the Periodic Table [36]. All these results cannot be connected to mass effect since heavier masses of cations would tend to smaller frequencies, which is not the case. The above results indicate that the stretching force constants in $BO_4$ tetrahedral molecules are dependent on the oxidation state of the B cation and consequently on the charge density of the $BO_4$ molecule, which is affected by the compound ionicity. The increase of the stretching force constant with increasing the oxidation state of the B cation (or with a higher charge density in the $BO_4$ tetrahedra) is due to the higher degree of $\sigma+\pi$ bonding between the B cation and the O anion present for the higher oxidation states (or for higher charge densities) [37]. In particular, in scheelite tungstates the Raman and IR $\nu_1$ and $\nu_3$ frequencies are different for each compound and the compound with the largest ionicity ($BaWO_4$); i.e., with the largest charge transfer from the $A^{2+}$ cation to the $WO_4^{2-}$ anion, gives the largest $\nu_1$ and $\nu_3$ frequencies. Note that the $\nu_3$ frequencies almost equal in all three alkaline-earth tungstates. On the other hand, the smaller $\nu_1$ and $\nu_3$ frequencies in PbWO$_4$, as compared to CaWO$_4$, SrWO$_4$ and BaWO$_4$, are likely due to the more soften W-O bond in PbWO$_4$ since the W-O distances are pretty similar in the four tungstates **[11,12]**. In summary, we think that the stretching force constant in scheelite tungstates depends on the charge density of the WO$_4$ molecule which is affected by the electronegativity of the A cation.

The WO$_4$ molecule has not yet been isolated in nature, therefore the internal modes A$_1$, E, and 2F$_2$ of the WO$_4$ molecule, usually named as $\nu_1$, $\nu_2$, $\nu_3$ and $\nu_4$ **[24]**, are not known. Raman measurements of tungstates solved in water give the following frequencies for the quasi-free WO$_4$ molecule **[30,35,36]**:

$$\nu_1 = 931 \text{ cm}^{-1} \qquad \nu_2 = 325 \text{ cm}^{-1}$$
$$\nu_3 = 838 \text{ cm}^{-1} \qquad \nu_4 = 325 \text{ cm}^{-1} \qquad (2)$$

However, the proven dependence of the stretching frequencies of the WO$_4$ molecule on the ionicity of the material and the fact that H$_2$O is a polar solvent that can transfer charge to the WO$_4^{2-}$ molecules cast a reasonable doubt about the validity of the above values as those corresponding to the quasi-free WO$_4$ molecule. In this sense, we must note that the stretching $\nu_1$ frequency for the quasi-free WO$_4$ molecule measured in water (931 cm$^{-1}$) is even higher than that measured in BaWO$_4$ (926 cm$^{-1}$). Therefore, we conclude that the most approximate frequencies for the free WO$_4$ molecule, neglecting the distortion of the WO$_4$ tetrahedra in the solid compounds, are those found in PbWO$_4$, which is the tungstate compound with smaller ionicity. In summary, we propose that the Raman modes of the isolated WO$_4$ tetrahedra must be rather similar to the following ones:

$$\nu_1 = 906 \text{ cm}^{-1} \qquad \nu_2 = 325 \text{ cm}^{-1}$$
$$\nu_3 = 760 \text{ cm}^{-1} \qquad \nu_4 = 360 \text{ cm}^{-1} \qquad (3)$$

The same reasoning can be applied to other scheelite compounds, like scheelite molybdates, and conclude that the most approximate frequencies for the free $MoO_4$ molecule are those found in scheelite-$PbMoO_4$. The small Davydov splittings (or factor group splittings) of the internal modes of the $WO_4$ molecule in stolzite give support for the assignment of the average frequencies of stolzite to the quasi-free $WO_4$ molecule. Even though in the free $WO_4$ molecule there should be no Davydov splitting of the $\nu_2$, $\nu_3$, and $\nu_4$ modes due to the interaction of equivalent interacting $WO_4$ molecules inside a unit cell, stolzite exhibits the smallest splittings of the known scheelite tungstates. In stolzite the Raman $\nu_2$ splitting is 5 cm$^{-1}$, the $\nu_3$ splitting is 14 cm$^{-1}$, and the $\nu_4$ splitting is 5 cm$^{-1}$ (see **Table I**). As regards the IR splittings, the $\nu_3$ splitting is about 8 cm$^{-1}$, and the $\nu_4$ splitting is 28 cm$^{-1}$ (see **Table II**). These splittings are considerably smaller than those in the scheelite alkaline-earth tungstates **[10]**. In particular, the $\nu_3$ splitting in the scheelite structure comes from the non-equivalence of the interactions of the four W-O bonds in the distorted tetrahedron. Therefore, the small $\nu_3$ splitting in $PbWO_4$ is indeed indicative of the small distortion of the $WO_4$ tetrahedron, as compared to that in scheelite alkaline-earth tungstates, as should be expected in a quasi-free $WO_4$ molecule.

Finally, to close this section we want to point out that we have observed several faint modes in the Raman spectrum at low pressures that are located at 621.4 cm$^{-1}$ and at 666.5 cm$^{-1}$ at 1 atm (see arrows in **Figs. 1 and 4**). These modes have frequency pressure coefficients of 1.0 and 2.3 cm$^{-1}$GPa$^{-1}$, respectively and their behaviors under pressure are displayed with open circles in **Fig. 3**. We have attributed this two Raman modes to the symmetric and asymmetric stretching modes of the silicon oil pressure medium since $Si(-O-CH_3)_n$ and $Si-CH_3$ chains of dimethylsiloxanes and trimethylsiloxanes in silicon oils are in the frequency region between 600 and 750 cm$^{-1}$ **[38,39,40]**.

**B. High-pressure phases.**

**Figure 4** shows the Raman spectra of the high-pressure phases of PbWO$_4$ at different pressures ranging from 6.2 to 17 GPa. At 6.2 GPa many Raman modes of the scheelite phase can still be observed together with new peaks that do not correspond to the scheelite phase (see exclamation marks in **Fig. 4**). Therefore, we have taken this value as the pressure of the onset of the first phase transition (see dotted line in **Fig. 3**). Apart from the new peaks appearing at 6.2 GPa, there are new Raman peaks appearing at 7.9 GPa (see asterisks in **Fig. 4**). In the following we will show that this pressure marks the onset of the second phase transition (see also dotted line in **Fig. 3**). The Raman modes of these two new phases coexist with those of the scheelite phase up to 9.0 GPa since the strongest peak of the scheelite phase ($\nu_1$(A$_g$) mode) is observed up to this pressure as a high-frequency shoulder of other high-frequency peaks. We must note that on decreasing pressure from 17 GPa the scheelite phase of PbWO$_4$ was recovered below 5 GPa.

The Raman spectra of PbWO$_4$ above 7.9 GPa are completely different from that of the scheelite phase, and are dominated by three high-frequency peaks located just at lower frequencies than that of the scheelite $\nu_1$(A$_g$) mode. The most striking features of the spectrum at 9.0 GPa are: 1) the large number of modes observed, which is even larger than the number expected for the fergusonite phase, as we discuss below; 2) some of the new modes are located in the phonon gap of the scheelite phase between 400 and 750 cm$^{-1}$; 3) the appearance of three modes near the scheelite $\nu_1$(A$_g$) mode; 4) the splitting of the scheelite $\nu_3$ modes between 700 and 800 cm$^{-1}$; and 5) the appearance of several well-separated broad modes between 300 and 500 cm$^{-1}$.

In the following discussion we will show that the new Raman peaks appearing at 6.2 GPa correspond to the PbWO$_4$-III phase, while the new Raman modes appearing at 7.9 GPa correspond to the fergusonite phase. We will show that, similarly to BaWO$_4$ **[10]**, PbWO$_4$ suffers a scheelite-to-fergusonite and a scheelite-to-P2$_1$/$n$ phase transitions, and that there is a region of coexistence of the scheelite, fergusonite, and P2$_1$/$n$ phases. Following the method used for BaWO$_4$ **[10]**, the assignment of the Raman modes appearing above 6.2 GPa to the fergusonite and PbWO$_4$-III phases is based on the classification of these modes into two types of modes: 1) modes that appear above 6.2 GPa and decrease in intensity above 10 GPa, but can be followed in pressure up to almost 14.6 or 16.7 GPa; 2) modes that appear above 7.9 GPa, that attain a maximum intensity around 9.0 GPa, and weaken more rapidly than the first ones above 10 GPa disappearing above 14 GPa. The different pressure behavior of these two types of modes leads us to believe that between 7.9 and 9.0 GPa we have a mixture of two high-pressure phases with the scheelite one and that between 9.0 and 14 GPa we have a coexistence of the two high-pressure phases. The assignment of the different Raman modes to the two high-pressure phases was difficult because the high-pressure Raman spectra of PbWO$_4$ above 6.2 GPa are considerably different from those reported for the alkaline-earth tungstates **[8,10,27,28]**. Fortunately, the high-frequency region of the Raman spectrum of PbWO$_4$ between 7.9 and 9.0 GPa resembles that of the Raman spectrum of BaWO$_4$ between 7.5 and 9.0 GPa **[10]**; so the different pressure behavior of the two types of new modes can be clearly seen in the three strong high-frequency stretching modes appearing in the Raman spectra between 850 and 950 cm$^{-1}$ above 7.9 GPa (see **Fig. 4**), as it was already observed in BaWO$_4$ **[10]**. In the spectrum at 7.9 GPa there are four strong high-frequency modes. The mode with highest frequency and intensity at 7.9 GPa is the scheelite $\nu_1$(A$_g$) mode, which is observed up to 9 GPa. However, above 9 GPa there are only three intense high-frequency modes that, on the light of their similar intensity, we think that derive from the scheelite $\nu_1$(A$_g$) mode **[10]**.

The mode with lowest frequency around 870 cm$^{-1}$ at 9 GPa decreases very rapidly in intensity with increasing pressure, whereas the other two high-frequency modes near 890 cm$^{-1}$ at 9 GPa decrease more slowly in intensity and remain clearly visible up to the highest pressure attained in our experiment. Therefore, on the light of the unstability of the ferrusonite structure versus the PbWO$_4$-III structure suggested by *ab initio* calculations **[12]**, we attribute the strong mode near 870 cm$^{-1}$ to the ferrusonite phase while the other two strong modes are assigned to the PbWO$_4$-III phase. A decrease in intensity above 10 GPa can be also observed in other modes like the two ferrusonite modes with lower frequencies near 40 and 50 cm$^{-1}$ (see asterisks in **Fig. 4**). In sections B.1 and B.2 we will discuss more deeply the nature of the different modes assigned to the high-pressure phases.

**B.1. Ferrusonite structure**

In **Ref. 10** it was noted that the Raman modes in CaWO$_4$ and SrWO$_4$ reported between 12 and 20 GPa **[27,28]** and the Raman modes in BaWO$_4$ appearing at 7.5 GPa and disappearing at 9 GPa **[10]** correspond to the ferrusonite phase. Besides, there is evidence of an experimental scheelite-to-ferrusonite phase transition in PbWO$_4$ **[12]**, Therefore, let us begin the study of the Raman peaks of the high-pressure phases in PbWO$_4$ by comparing the expected ferrusonite modes with those reported previously **[10,27,28]**. The centrosymmetric ferrusonite structure (I2/*a*, SG No. 15, Z = 4) should have 36 vibrational modes at the zone centre, like in BaWO$_4$ **[10]**, with the following mechanical representation:

$$\Gamma = 8A_g + 8A_u + 10B_g + 10B_u \qquad (4)$$

The 18 gerade (g) modes are Raman active and the 18 ungerade (u) modes are IR active. The 18 Raman-active modes derive from the reduction of the tetragonal C$_{4h}$ symmetry of the scheelite structure to the monoclinic C$_{2h}$ symmetry of the ferrusonite structure. In particular, every A$_g$ and every B$_g$ scheelite mode transforms into an A$_g$ mode of the monoclinic symmetry, while every doubly degenerate $E_g$ scheelite mode transforms into two $B_g$ modes of the monoclinic symmetry. Likewise to $BaWO_4$ **[10]**, many Raman modes of the scheelite structure have weakened considerably or disappeared at 9.0 GPa and the number of new modes measured at 9.0 GPa exceed the number of modes expected for the fergusonite structure. These results point out that either the high-pressure phase is not fergusonite or that there is a mixture of phases with one phase being fergusonite, as evidenced by ADXRD measurements **[12]**. The above classification of the two classes of Raman modes appearing above 6.2 GPa and 7.9 GPa makes clear that there is a mixture of phases between 7.9 and 14 GPa. Therefore, taking into account first of all that recent ADXRD measurements observed a fergusonite phase above 9 GPa **[12]**, that the results of *ab initio* total-energy calculations showed the larger stability of the $PbWO_4$-III phase respect to the fergusonite phase at high pressures **[12]**, and finally that the number of modes appearing at 7.9 GPa and disappearing near 13 GPa is around sixteen, we attribute the modes that appear above 7.9 GPa and disappear progressively above 10 GPa to the fergusonite structure.

**Figs. 5, 6 and 7** show enlarged portions of the high-pressure Raman spectra of $PbWO_4$ for a detailed analysis of the behavior of this compound under pressure. The Raman modes assigned to the fergusonite phase are marked with asterisks and the calculated frequencies of the eighteen fergusonite Raman modes at 9 GPa are indicated at the bottom of the figures. In these figures, the appearance of the modes assigned to the fergusonite phase at 7.9 GPa and their fading beginning at 10 GPa can be seen more clearly than in **Fig. 4**. **Fig. 3** shows the pressure dependence of the frequencies of the fergusonite Raman modes (empty squares) in $PbWO_4$. **Table III** summarizes the experimental and theoretical Raman frequencies, and their pressure coefficients for the fergusonite modes at 9 GPa. For completeness, we also report in **Table IV** the frequencies and pressure coefficients of the calculated IR-active modes of fergusonite-

PbWO$_4$ at 9 GPa. A major conclusion drawn from our calculations and seen in **Figs. 6 and 7** is that there is a phonon gap between 470 and 700 cm$^{-1}$ in ferguson ite-PbWO$_4$ similar to the one observed in scheelite-PbWO$_4$. This result is similar to that found in BaWO$_4$ **[10]** and confirms that the fergusonite phase retains at least partially the tetrahedral W coordination of the scheelite phase (see discussion in section C). It also leads us to conclude that the Raman modes observed at high pressures in this phonon gap belong to a phase other than the scheelite or the fergusonite.

It is difficult to determine the frequency and symmetry of all the modes in the high-pressure phases because the small difference in the rate of intensity decrease between the Raman modes of the fergusonite and PbWO$_4$-III phases makes difficult the assignment of some modes to either of both structures. For the identification of the fergusonite Raman modes in PbWO$_4$ we can make use of the similarities between the scheelite and fergusonite structures already noted in **Ref. 10**, and the aid of our lattice dynamics calculations. As already commented, the most clear fergusonite mode is the mode located at 870 cm$^{-1}$ at 9 GPa and corresponding to the highest stretching A$_g$ mode arising from the scheelite $\nu_1$(A$_g$) mode. Calculations at 9 GPa locate this mode around 846 cm$^{-1}$; i.e., with an error smaller than 3%, and clearly at smaller frequencies than the other two strong Raman peaks. A striking difference between the scheelite-to-fergusonite phase transitions in PbWO$_4$ and alkaline-earth tungstates is the strong decrease in frequency between the scheelite $\nu_1$(A$_g$) mode and the related fergusonite A$_g$ mode in PbWO$_4$ (40 cm$^{-1}$), as compared to the slight change in frequency observed in the alkaline-earth tungstates (5 cm$^{-1}$) **[10,27,28]**. Curiously enough, on the basis of the frequencies reported by Jayaraman *et al.* **[9]** the jump observed in PbWO$_4$ seems not occur in PbMoO$_4$. We will comment this fact in section C when discussing the W coordination in the high-pressure phases. The other three high-frequency stretching

ferguson­ite modes can be located and assigned with the aid of their calculated frequencies (also slightly underestimated) and pressure coefficients, and due to its larger intensity in the Raman spectrum at 9 GPa (see **Fig. 7** and **Table III**). Note that the fergusonite phase is dominant in the Raman spectrum of $PbWO_4$ at 9 GPa in agreement with ADXRD measurements **[12]**.

Concerning the five fergusonite modes arising from the four scheelite $\nu_2$ and $\nu_4$ modes, we have tentatively assigned the 446 and 471 $cm^{-1}$ experimental peaks to the two fergusonite $B_g$ modes arising from the scheelite $\nu_4(E_g)$ mode. Support for this assignment is given by the closeness of the calculated modes (440 and 467 $cm^{-1}$ at 9 GPa) and by the qualitative agreement between the large pressure coefficients measured and calculated for these two modes (see **Table III**). In section C we will discuss why the pressure coefficients of these two fergusonite modes are considerably larger than those of other fergusonite modes. The mode measured at 396 $cm^{-1}$ at 9 GPa has been identified as the fergusonite $A_g$ mode arising from the scheelite $\nu_4(A_g)$ mode with a calculated frequency of 377 $cm^{-1}$ at 9 GPa. And finally, the two modes located at 320 and 352 $cm^{-1}$ at 9 GPa have been assigned to the fergusonite $A_g$ modes arising from the scheelite $\nu_2$ modes with calculated frequencies of 303 and 365 $cm^{-1}$ at 9 GPa. The three above assignments are supported also by the qualitative agreement between the measured and calculated pressure coefficients. We should note that even though the measured pressure coefficients are smaller than the calculated ones for the fergusonite modes, in general the qualitative agreement guiding our assignments is good (see **Table III**).

As regards to the external fergusonite modes, first of all we have to note that contrary to what happens in scheelite alkaline-earth tungstates, the translational (T) modes in stolzite are well separated in frequency from the rotational (R) modes.

Therefore, we expect six Raman modes below 200 cm$^{-1}$ arising from the four T modes plus three Raman modes above 200 cm$^{-1}$ arising from the two R modes in fergusonite-PbWO$_4$, as indeed observed from our calculations in **Figs. 5 and 6**. However, like in alkaline-earth tungstates **[10]**, we have not found experimentally the two fergusonite modes coming from the scheelite R(E$_g$) mode in PbWO$_4$. We believe that their absence can be due to the weak intensity of these two modes already in the scheelite phase. Furthermore, these two fergusonite modes have very large calculated pressure coefficients that do not agree with none of our observed modes. However, we think that we have experimentally found the fergusonite A$_g$ mode deriving from the scheelite R(A$_g$) mode near 250 cm$^{-1}$ at 9 GPa on the basis of the measured and calculated frequency and pressure coefficient.

In order to identify the fergusonite modes arising from the scheelite T modes, we will begin discussing those with smaller frequencies. There are two modes with frequencies near 44 and 53 cm$^{-1}$ at 9 GPa (see asterisks in **Fig. 5**) whose intensity decrease strongly above 10 GPa. We have tentatively attributed the mode at 53 cm$^{-1}$ to the lowermost fergusonite A$_g$ mode arising from the lowermost scheelite T(B$_g$) mode. On the other hand, we have attributed the mode at 44 cm$^{-1}$ to the lowermost fergusonite B$_g$ mode arising from the lowermost scheelite T(E$_g$) mode. The other fergusonite B$_g$ mode arising from this scheelite T mode is located around 63 cm$^{-1}$ in agreement with *ab initio* calculations. The present assignment for the two lowermost fergusonite modes is supported by the small pressure coefficient of the fergusonite mode derived from the soft scheelite T(B$_g$) mode with negative pressure coefficient. However, the calculated fergusonite frequencies for these two modes are considerably higher than the experimentally observed modes (see **Table III**). A similar overestimation of the experimental frequencies was also found for the external modes with the smallest frequencies in fergusonite-BaWO$_4$ **[10]**. It is curious that, unlike BaWO$_4$, the fergusonite mode arising from the soft scheelite T($B_g$) mode in PbWO$_4$ does not exhibit a negative pressure coefficient (0.4 cm$^{-1}$/GPa). We think that this could be due to the strong deformation of the WO$_4$ tetrahedra in the fergusonite phase of PbWO$_4$ which is shown by *ab initio* total-energy calculations **[12]** and discussed in section C. The deformation of the fergusonite structure in PbWO$_4$ makes this structure more stable in PbWO$_4$ than in BaWO$_4$, where the instability of this phase was evidenced by the quick disappearance of its Raman modes above 8.2 GPa **[10]**. In contrast, Raman modes of fergusonite-PbWO$_4$ last almost up to 14.6 GPa. Finally, there are three calculated fergusonite modes between 100 and 200 cm$^{-1}$ arising from the topmost T modes with similar pressure coefficients to those measured for the three experimental modes at 136, 144, 152 cm$^{-1}$. The agreement between experimental and the theoretical frequencies and pressure coefficients justifies the above assignments (see **Table III**).

### B.2. PbWO$_4$-III structure

The existence of a high-pressure structure in PbWO$_4$ having the PbWO$_4$-III structure **[7]** has been recently suggested by Errandonea *et al.* **[12]**. The experimental powder ADXRD patterns above 15 GPa are compatible with the PbWO$_4$-III structure **[12]**. In this section, we will show with the help of our *ab initio* lattice dynamics calculations that the Raman modes of PbWO$_4$ appearing above 6.2 GPa; i.e. before the appearance of modes corresponding to the fergusonite phase at 7.9 GPa, are consistent with the *ab initio* total-energy calculations, which yield a lower value (5.3 GPa) for the scheelite/PbWO$_4$-III coexistence pressure than for the scheelite/fergusonite one (8 GPa) **[12]**. This result lead us to attribute the new Raman modes appearing at 6.2 GPa and lasting up to 17 GPa to the PbWO$_4$-III phase (see exclamation signs in **Figs. 5, 6 and 7**). This phase shows a distorted octahedral W-O coordination which can explain the Raman modes observed above 6.2 GPa in the phonon gap of the scheelite and ferguson ite phases. Similar modes have also been observed in BaWO$_4$ above 6.9 GPa **[10]** and in BaMoO$_4$ above 9 GPa **[41,42]**.

The centrosymmetric PbWO$_4$-III structure is isomorphous to the BaWO$_4$-II structure (P2$_1$/*n*, SG No. 14, Z = 8) **[7, 10]** and group theoretical considerations lead to the following mechanical representation of vibrational modes at Γ **[22]**:

$$\Gamma = 36A_g + 36 A_u + 36B_g + 36B_u \qquad (5)$$

There are 72 Raman-active (g) modes and 72 IR-active (u) modes, of which one A$_u$ and two B$_u$ are the acoustic modes. One thus expects four times more Raman modes in the PbWO$_4$-III phase than in the ferguson ite phase. The experimental assignment of the mode symmetry in the PbWO$_4$-III phase is difficult because of its mixture with the ferguson ite phase, because of the impossibility of testing the polarization selection rules of the Raman modes inside the DAC, and because the number of modes that can be clearly resolved in the experimental Raman spectra above 6.2 GPa is around 37; i.e, about half the number of expected modes for the PbWO$_4$-III phase.

In **Ref. 10** we listed a number of factors that can explain why a large number of modes of the PbWO$_4$-III structure are not observed experimentally. In any case, the observation of most of the *ab initio* predicted modes for the PbWO$_4$-III phase in certain frequency ranges, especially in the phonon gap of the scheelite and ferguson ite phases, evidences that there is no other structure with higher symmetry, like LaTaO$_4$ (P2$_1$/*c*, SG No. 14, Z = 4) or raspite with half the number of expected modes than the PbWO$_4$-III phase, which could give account for the experimentally observed modes.

The pressure dependence of the Raman modes observed experimentally and attributed to the PbWO$_4$-III phase is shown in **Fig. 3** with solid triangles. The experimentally observed frequencies and pressure coefficients of the Raman modes attributed to the PbWO$_4$-III phase at 13.7 GPa are summarized in **Table V**. The assignment of the modes corresponding to the PbWO$_4$-III has been done with the help of lattice dynamics calculations for this phase at 9 GPa. **Tables VI** and **VII** summarize the calculated frequencies and symmetries of the Raman and IR modes attributed to the PbWO$_4$-III phase at 9 GPa, respectively. The frequencies of the Raman modes of the PbWO$_4$-III phase calculated at 9 GPa are marked at the bottom of **Figs. 5, 6 and 7**.

As previously commented, our assignment of the high-pressure phase of PbWO$_4$, stable between 6.2 GPa and 17 GPa, to the PbWO$_4$-III phase is supported by the observation, in several cases, of the four peaks expected for each fergusonite mode at frequencies close to those obtained from our lattice dynamics calculations. Clear examples of this fact can be seen in the high-frequency region above 500 cm$^{-1}$, due to the smaller density of modes in this region. The Raman spectra above 9 GPa show two strong high-frequency modes, located at 899 and 914 cm$^{-1}$ at 13.7 GPa, in addition to two weak high-frequency modes, located at 935 and 949 cm$^{-1}$ at 13.7 GPa, that appear as shoulders rather than as full peaks (see exclamation marks in **Fig. 7**). *Ab initio* calculations for the PbWO$_4$-III phase at 9 GPa show that many modes in this phase are grouped in pairs (see marks at the bottom of **Figs. 5, 6, and 7**). In particular, there are two groups of high-frequency modes located around 860 and 910 cm$^{-1}$ at 9 GPa. Therefore, we attribute these four modes to phonons in the PbWO$_4$-III phase that likely arise from the splitting of the scheelite $\nu_1$(A$_g$) mode or its related fergusonite A$_g$ mode despite the calculated frequencies are around 30 cm$^{-1}$ (4%) below our experimentally observed modes. This is the clearest example we have found of the observation of four modes in the PbWO$_4$-III phase for every fergusonite mode, supporting the PbWO$_4$-III nature of the high-pressure phase.

*Ab initio* calculations show four fergusonite modes and sixteen PbWO$_4$-III modes above 550 cm$^{-1}$ (see **Fig. 7**), in agreement with the 4:1 PbWO$_4$-III/fergusonite mode ratio previously commented. Therefore, we think that many of the Raman modes of the PbWO$_4$-III phase can be reasonably identified at least in the high-frequency region with smaller density of Raman peaks. In this sense, the two modes of the calculated pairs can be experimentally observed in many cases above 500 cm$^{-1}$. For instance, there are two pairs of calculated modes around 500 and other two around 575 cm$^{-1}$ at 9 GPa (see marks at the bottom in **Figs. 6 and 7**). One can distinguish these two pairs of modes between 520 and 540 cm$^{-1}$ and between 575 and 600 cm$^{-1}$ at 9 GPa (see exclamation marks in **Figs. 6 and 7**). In a similar way, there is a pair of calculated modes near 640 cm$^{-1}$ that can be correlated to the broad band located 670 cm$^{-1}$ at 9 GPa. Furthermore, we believe that the two modes observed at 693 and 710 cm$^{-1}$ at 13.7 GPa correspond to the calculated pair located near 690 cm$^{-1}$ at 9 GPa, and that the three calculated modes around 925 cm$^{-1}$ at 9 GPa cannot be clearly observed but contribute to the broad band marked with an interrogation mark in the Raman spectra of **Fig. 7** above 12.6 GPa. Finally, we tentatively attribute the modes located at 762, 790, and 820 cm$^{-1}$ at 9 GPa to those calculated at 761, 774, and 802 cm$^{-1}$ at 9 GPa. All the above examples give evidence that the PbWO$_4$-III phase can be reasonably identified, despite the number of Raman modes distinguished in the experimental spectra at high pressures is well below the number of expected modes for this phase. We are not going to discuss the correlation between experimentally observed and calculated modes of the PbWO$_4$-III phase in the low-frequency region because it is much more complicated due to the overlapping of modes in this region with a higher density of Raman modes.

Finally, we should mention that the coexistence of the scheelite, fergusonite and PbWO$_4$-III phases is possible due to the kinetic hindrance of the reconstructive scheelite-to-PbWO$_4$-III transition and the displacive second-order nature of the scheelite-to-fergusonite transition **[43]**. This is the same case as in BaWO$_4$ **[10]**. The kinetic hindrance of the scheelite-to-PbWO$_4$-III transition can be clearly observed in our Raman spectra between 6.9 and 10 GPa by checking the intensity of the two strong high-frequency peaks attributed to the PbWO$_4$-III phase and located at 899 and 914 cm$^{-1}$ at 13.7 GPa. The low-frequency mode of this pair grows in intensity between 9 and 12.6 GPa becoming as intense as the high-frequency mode of this pair only at 14.6 GPa. Besides, it can be observed an overall change in the Raman spectrum between 9 and 12.6 GPa with a decrease in intensity in many fergusonite modes and the appearance of some new modes at 12.6 and 13.7 GPa. All these results indicate that the scheelite-to-PbWO$_4$-III phase transition is not completed up to 14.6 GPa (see dashed line in **Fig. 3**). This is in good agreement with the results of ADXRD and XANES that showed a different phase from fergusonite above 15.6 and 16.7 GPa, respectively **[12]**. We believe that the Raman spectrum above 13.7 GPa corresponds mainly to the PbWO$_4$-III phase with the stretching fergusonite A$_g$ mode testifying the presence of a vanishing fergusonite phase (see **Fig. 7**). The presence of some modes in the PbWO$_4$–III phase with zero or even negative pressure coefficient, the overall decrease in intensity of the spectrum above 14.6 GPa, and the lack of good ADXRD spectra even at 10 GPa **[12]**, lead us to suspect that the PbWO$_4$–III phase is not very stable and could tend to another phase or to amorphisation above 17 GPa. However, on downstroke from 17 GPa the appearance of the strongest Raman peak of the scheelite phase below 5 GPa after a considerable hysteresis indicates that the scheelite-to-fergusonite and scheelite-to-PbWO$_4$-III phase transitions are reversible, as in the case of BaWO$_4$ **[10]**. This result is in agreement with **Refs. 9 and 12**.

In summary, we can conclude that the onset of the scheelite-to-PbWO$_4$-III phase transition is found around 6.2 GPa, being followed by a scheelite-to-fergusonite phase transition around 7.9 GPa. These phase transition pressures are in excellent agreement with the pressure of the scheelite-to-PbWO$_4$-III transition (5.3 GPa) and the pressure of the scheelite-to-fergusonite transition (8.0 GPa) found recently by *ab initio* total-energy calculations **[12]**. The observation of the coexistence of the scheelite, fergusonite, and PbWO$_4$-III phases between 7.9 GPa and 9.0 GPa, and the coexistence of the fergusonite and PbWO$_4$-III phases up to 14.6 GPa, can only be explained by the hindrance of the reconstructive scheelite-to-PbWO$_4$-III phase transition that favors the observation of the second-order scheelite-to-fergusonite phase transition, as already discussed in **Refs. 10, 12 and 43**.

We want to close this section showing that our results for the fergusonite and PbWO$_4$-III phases can also give account for the results of PbWO$_4$ reported up to 9 GPa by Jayaraman *et al*. **[9]** in the same way as in section B.1 we showed that our results for stolzite are consistent with those previously reported **[9]**. In **Ref. 9**, the three high-frequency stretching modes arising from the scheelite $\nu_1$(A$_g$) mode were observed above 4.5 GPa **[9]**. As regards to the modes of the PbWO$_4$-III phase, one of the two high-frequency stretching modes assigned to the PbWO$_4$-III structure near 900 cm$^{-1}$ is reported at 4.5 GPa (2.4 GPa below our measurements), and the other one above 6.5 GPa, likely due to their overlapping and limited resolution. Besides, two modes near 300 cm$^{-1}$, and two modes around 800 cm$^{-1}$ were observed at 4.5 GPa in agreement with our assignment of these modes to the PbWO$_4$-III phase. Regarding the fergusonite modes, the high-frequency A$_g$ stretching mode at 870 cm$^{-1}$ is in fact reported by Jayaraman *et al*. above 6.5 GPa (1.4 GPa below our measurements). The major disagreement between the present and previous works seems to be the assignment of the lowest-frequency mode around 45 cm$^{-1}$. This mode was previously observed above 4.5 GPa **[9]**, but we have observed it above 9 GPa. The observation of this mode already at 4.5 GPa would suggest that this mode could correspond to the PbWO$_4$-III phase and not to the fergusonite phase, as we have considered. We can not rule out the possibility that this mode corresponds to the PbWO$_4$-III phase, given the frequency difference between the measured and calculated fergusonite mode already commented, and the fact that our *ab initio* calculations locate two modes of the PbWO$_4$-III phase near 44 cm$^{-1}$ at 9 GPa.

Finally, we have to point out that, in our opinion, the Raman spectra of PbWO$_4$ in **Ref. 9** agree qualitatively with ours despite we observe the phase transitions at rather higher pressures than in **Ref. 9**. Furthermore, we think that the coexistence of the two high-pressure phases is also observed in **Ref. 9** between 6.5 and 8.5 GPa.

**C. Tungsten coordination in high-pressure phases**

Our assignment of the Raman modes in the high-pressure phases to the fergusonite and PbWO$_4$-III phases is coherent with the change of W coordination from tetrahedral to octahedral with increasing pressure. Features that support a basically tetrahedral W coordination in the fergusonite phase and an octahedral W coordination in the PbWO$_4$–III phase are: 1) the presence of a phonon gap in the fergusonite phase (between 470 and 700 cm$^{-1}$) similar to that of the scheelite phase; and 2) the appearance of modes of the PbWO$_4$-III phase in the phonon gap of the scheelite and fergusonite phases. However, the tetrahedral W coordination in the fergusonite phase is not fully compatible with the strong decrease of the highest-frequency stretching A$_g$ mode from 912 cm$^{-1}$ in the scheelite phase to 870 cm$^{-1}$ in the fergusonite phase. This strong decrease in frequency suggests larger W-O distances in the fergusonite phase than in the scheelite phase compatible with an octahedral W coordination in the high-pressure phase. Furthermore, in alkaline-earth tungstates, where the fergusonite phase retains the tetrahedral W coordination of the scheelite phase, there is a very small change in the high-frequency A$_g$ stretching mode at the scheelite-to-fergusonite transition due to the small change of the shorter W-O bond distance between the two phases **[10]**. In the discussion below we will show that the large decrease of the highest frequency mode of the fergusonite phase in PbWO$_4$ is clearly due to the tendency of fergusonite-PbWO$_4$ to octahedral W coordination.

As already commented in **Ref. 10**, there is a relationship between the frequencies of the stretching W-O modes and the bond distance R (in Å) between W and O in tungsten oxides **[44]**,

$$\omega \text{ (cm}^{-1}) = 25823 \exp(-1.902 \cdot R) \tag{6}$$

and there is a relationship between the bond distance R (in Å) and the Pauling's bond strength s, which for tungsten oxides is **[45]**:

$$s_{W-O} = (R/1.904)^{-6} \tag{7}$$

with $s_{W-O}$ given in valence units (v.u.) and with 1.904 Å being the bond distance corresponding to the unit valence.

By taking the values of 752, 766, and 906 cm$^{-1}$ as the stretching frequencies of stolzite at 1 atm, we can estimate with **Eqs. (6) and (7)** three W-O bond distances 1.86, 1.85, and 1.76 Å, and three bond strengths 1.15, 1.19, and 1.59 v.u., respectively. With the three bond strengths we can estimate a total valence of 5.53 v.u. which matches approximately the formal valence of the W$^{6+}$ ion (6) **[44]**. To obtain this result, we must consider a double contribution of the shortest bond distance on the basis of the fourfold W coordination (four W-O distances instead of three) in the scheelite structure **[44]**. In this configuration, the double contribution of the shorter distance leads to an average W-O distance of 1.808 Å in good agreement with the estimated W-O bond distance from x-ray and neutron diffraction measurements **[12,14]**. The above result agrees also with the expected ideal W-O bond distance in tetrahedral W coordination (1.78 Å), which corresponds to a Pauling's bond strength of 1.5 v.u. in **Eq. (7)** for each of the four W-O bonds **[44]**. However, the rather small value for the total W$^{6+}$ ion valence in stolzite (5.5 v.u.) than in alkaline-earth tungstates (5.8-5-9 v.u.) **[10]** using empirical data and **Eqs. (6) and (7)** could suggest a small W-O bond strength in stolzite, responsible for the smaller frequencies of the internal modes of the WO$_4$ tetrahedron, and consequently a tendency of stolzite to octahedral W coordination. Note that Hg and Pb have similar masses and electronegativities and HgWO$_4$ crystallizes in the C2/*c* structure where W is octahedrally coordinated **[46]**.

In a similar way, we can take the values of 872, 779, 725 and 693 cm$^{-1}$ as the stretching frequencies of fergusonite-PbWO$_4$ at 9 GPa (see **table III**). With these four frequencies we have estimated the W-O bond distances of 1.78, 1.84, 1.88, and 1.90 Å, and the strengths of 1.49, 1.23, 1.1 and 1.0 v.u., respectively. In total they sum 4.8 v.u. for the W ion, i.e., much lower than the formal valence of the W$^{6+}$ ion (6). This result casts also a doubt about the fourfold coordination in fergusonite-PbWO$_4$, and suggests a higher W coordination in this phase. If this were the case, additional stretching modes should be considered in the calculations using **Eqs. (6) and (7)**. Taking the next two internal modes in order of decreasing frequency; i.e., those at 471 and 446 cm$^{-1}$ at 9 GPa, as the remaining stretching modes in fergusonite-PbWO$_4$, we found that they would correspond to W-O bond distances of 2.10 and 2.13 Å, respectively according to **Eq. (6)**, and they would yield bond strengths of 0.55 and 0.51 v.u., according to **Eq. (7)**. Therefore, adding these values to the previous results we would get a total sum of 5.86 v.u., that matches nicely the formal valence of the W$^{6+}$ ion. This result clearly indicates that the fergusonite structure in PbWO$_4$ shows a 4+2 coordination for W, and that this is the reason for the strong decrease of the high-frequency A$_g$ stretching mode arising from the scheelite $\nu_1$(A$_g$) mode after the scheelite-to-fergusonite phase transition. This strong decrease of the highest stretching frequency must be related to a structural change in the W-O distances in PbWO$_4$ giving a strong distortion of the WO$_4$ tetrahedra after the scheelite-to-fergusonite transition.

**Figure 8** shows *ab initio* calculations of the W-O bond distances in the scheelite and fergusonite phases of PbWO$_4$. Details of these calculations are given in **Ref. 12**. It can be seen how the two different W-O distances in the scheelite phase change considerably after the phase transition. The four shorter and equal W-O bond distances in the scheelite phase split into two pairs of different and slightly larger distances. On the other hand, the four larger and equal second-neighbor W-O distances in the scheelite phase suffer a big splitting at the phase transition leading to two W-O bond distances around 2.25 Å at 9.5 GPa. This big decrease of second-neighbor W-O distances justify the 4+2 coordination for W in the ferguson phase of $PbWO_4$. Moreover, the big change of interatomic distances and the distortion of the W-O bonds in ferguson-$PbWO_4$ are evidences that this phase in $PbWO_4$ acts as a bridge phase between a structure with fourfold and another with sixfold W coordination. Furthermore, we should note that the estimated W-O bond distances using **Eq. (6)** and corresponding to the above 6 ferguson stretching Raman modes taken in pairs (1.74+1.84, 1.88+1.90, and 2.10+2.13), give average W-O bond distances of 1.79, 1.89, and 2.12 Å at 9 GPa. These values are reasonably close to those found by *ab initio* calculations in **Fig. 8** for the ferguson structure at 9.5 GPa (1.80, 1.88, and 2.25 Å), and confirm that **Eqs. (6) and (7)** allow the prediction of the 4+2 coordination for W in the ferguson structure of $PbWO_4$. Finally, we want to point out that the change of W coordination after the scheelite-to-ferguson transition in $PbWO_4$ unlike in alkaline-earth tungstates and the sensitivity of XANES to coordination changes allows us to explain why XANES measurements detected the scheelite-to-ferguson phase transition in $PbWO_4$ around 9 GPa, but not in $BaWO_4$ **[12]**. Furthermore, the change from 4+2 coordination for W in ferguson-$PbWO_4$ to 6 coordination in the $PbWO_4$-III phase allows explaining the faint observation of the ferguson-to-$PbWO_4$-III phase transition in XANES at 16.7 GPa **[12]**. The lack of a clear transition between the ferguson and $PbWO_4$–III phases, as found in XANES measurements **[12]**, is coherent with the large monoclinic distortion of the ferguson structure in $PbWO_4$, that is reflected in the large distortion of the $WO_4$ tetrahedra in this phase. The larger distortion of the $WO_4$ tetrahedra in ferguson-$PbWO_4$, as compared to the ferguson phase of alkaline-earth tungstates, is opposite to that in scheelite-PbWO$_4$, which exhibits the least distorted WO$_4$ tetrahedra, as already discussed.

It must be noted that the theoretical knowledge of the behaviour W-O bond distances with pressure reported in **Fig. 8** for ferrusonite-PbWO$_4$ allows an estimation of the frequencies and pressure coefficients of the internal stretching modes of the tungstate group in ferrusonite-PbWO$_4$ by using **Eqs. (6) and (7)**. We are not going to do such a task for the sake of simplicity, but we want to stress that the two additional stretching ferrusonite modes observed experimentally at 446 and 471 cm$^{-1}$ at 9 GPa, and that *ab initio* calculations locate at 458 and 462 cm$^{-1}$ at 9 GPa, have calculated pressure coefficients considerably larger than the those of the other ferrusonite modes (see **Table III**). These high pressure coefficients can be understood if we consider that the estimated W-O bond distances associated to these two modes (2.10 and 2.13 Å at 9 GPa) correspond to the W-O bond distance theoretically calculated at 2.25 Å at 9.5 GPa (see **Fig. 8**). The large pressure coefficients of the above two modes are coherent with the large decrease of the associated W-O bond length under compression due to the distortion of the tungstate group observed in **Fig. 8**. Note that in BaWO$_4$ the corresponding ferrusonite modes (calculated to be at 362 and 363 cm$^{-1}$ at 7.5 GPa) have rather similar calculated pressure coefficients than the other ferrusonite modes **[10]**.

Another conclusion that can be drawn from the fact that the ferrusonite phase of PbWO$_4$ is a structure acting as a bridge between 4 and 6 coordination for W, is that maybe modes of the ferrusonite phase overlap with modes of the PbWO$_4$–III phase due to the small distortion between both phases. This fact makes even more difficult the assignment of modes of both phases. The above conclusion could be supported by: 1) the small fading rate of some ferrusonite modes above 10 GPa; 2) the appearance of a few PbWO$_4$–III modes above 12.6 GPa; and 3) the observation of some ferrusonite modes (much weaker in intensity than the mode at 870 cm$^{-1}$ at 9 GPa) up to pressures of 14.6 GPa (see **Figs. 5, 6, and 7**).

Finally, the application of Hardcastle and Wachs' rules to the PbWO$_4$-III phase can help in assigning the first-order modes of this low symmetry phase provided that the WO$_6$ octahedra can be regarded as almost independent units **[44]**. In the characterization of the PbWO$_4$-III structure at 1 atm, the following W-O distances were reported **[7]**: 1.76, 1.79, 1.81, 1.81, 1.83, 1.88, 1.91, 2.03, 2.10, 2.16, 2.17, and 2.26 Å. The average distances of **Ref. 7** at 1 atm taken in pairs are: 1.775, 1.81, 1.855, 1.97, 2.13, and 2.22 Å, and **Eq. (7)** allows us to obtain the following Pauling's bond strengths in valence units (v.u.): 1.75, 1.5, 1.23, 0.76, 0.41, 0.29 v.u. Altogether they sum 5.9 v.u., which fits nicely the formal valence of the W$^{6+}$ ion **[44]**. Therefore, this result clearly suggest that W has an octahedral coordination in the PbWO$_4$-III phase and that the W-O bond distances reported in **Ref. 7** could be used to estimate the stretching modes of the PbWO$_4$–III phase. With data of **Ref. 7** and **Eq. (6)** we have calculated the stretching frequencies: 908, 858, 826, 795, 723, 683, 543, 458, 424, 416, and 351 cm$^{-1}$. These values agree qualitatively with the average positions of the stretching modes calculated at 9 GPa by first principles and with our assignments of the PbWO$_4$-III Raman peaks (see **Tables IV and V**, and **Figs. 6 and 7**).

**D. Phase transitions in other related compounds**

As already commented in a previous work **[10],** the attribution of the Raman spectra of CaWO$_4$ and SrWO$_4$ above 10 GPa, and of BaWO$_4$ above 7.5 GPa, to the fergusonite phase is supported by the comparison of their Raman spectra with the Raman spectrum of HgWO$_4$ at 1 atm **[46]**. We can see that the Raman spectrum of HgWO$_4$ at 1 atm can be also compared to those of PbWO$_4$ above 8 GPa. HgWO$_4$ is a compound crystallizing also in a monoclinic structure (C2/c, SG No. 15, Z = 4) with octahedral coordination, but that can be regarded as tetrahedral since two W-O distances are far more large than the other four [46]. In this respect $HgWO_4$ would share characteristics of the fergusonite phases observed in $PbWO_4$ and alkaline earth tungstates. The three stretching modes with highest frequencies in $HgWO_4$ around 850, 815, and 930 cm$^{-1}$, are more similar to the fergusonite phases of alkaline-earth tungstates than to those in fergusonite-$PbWO_4$. However, there is a high-frequency stretching mode near 700 cm$^{-1}$ in $HgWO_4$ similar to the 693 cm$^{-1}$ mode in fergusonite-$PbWO_4$ at 9 GPa. As regards to the low-frequency modes in $HgWO_4$, there are a couple of modes near 500 cm$^{-1}$ that could be part of the three fergusonite modes coming from the scheelite $v_4(B_g+E_g)$ modes, in a similar way to the stretching modes found near 470 cm$^{-1}$ in fergusonite-$PbWO_4$. We think that the similar stretching mode frequencies around 500 cm$^{-1}$ and 700 cm$^{-1}$ in fergusonite-$PbWO_4$ and $HgWO_4$ are features related to the tendency of W to octahedral coordination in these two tungstates with more electronegative A cations like Hg and Pb, and where d orbitals can also play a significant role. In any case, new and more detailed RT Raman spectra of monoclinic fergusonite-like $HgWO_4$ would be of help in the definite assignment of all its modes, especially in the low frequency region.

We would like to highlight also that the pressure-driven transitions for scheelite $PbWO_4$ could be the same as those for scheelites $AgReO_4$ and $PbMoO_4$. This conclusion is obtained on the basis of the positions of the three compounds in Bastide's diagram [47] and on the similarities of their high-pressure Raman spectra. In the case of $AgReO_4$, we can compare it directly to $PbWO_4$ since Raman spectra up to 18 GPa were previously reported by J.W. Otto *et al*. [48]. However, in the case of $PbMoO_4$ the lack of detailed Raman spectrum at high pressures allows us to discuss it only on the basis of the reported frequencies of the high-pressure phases [9]. Both scheelite $AgReO_4$ and $PbMoO_4$ are located in south direction with respect to $PbWO_4$ in Bastide's diagram, therefore on the basis of the north-east rule for pressure increase in Bastide's diagram

**[47]** one can predict a larger stability of the scheelite phase in these compounds than in PbWO$_4$, as indeed observed. The scheelite phase in PbMoO$_4$ is found up to 9.5 GPa, and that of AgReO$_4$ is observed up to 11 GPa. We can interpret the Raman spectra of PbMoO$_4$ and AgReO$_4$ on the basis of: 1) the comparison of the scheelite frequencies and pressure coefficients of both compounds with those of PbWO$_4$, and 2) the comparison of the high-pressure spectrum of AgReO$_4$ at 14 GPa and the spectra of PbWO$_4$ between 9 and 10 GPa.

Regarding the scheelite phase, the spectra of the three compounds exhibit similar bands with similar frequencies and pressure dependences: 1) there is a gap between the external T and R modes, being this gap larger in PbWO$_4$ and PbMoO$_4$ than in AgReO$_4$; 2) the pressure coefficients of the scheelite stretching modes in these compounds do not scale with the length of the c-axis, as it occurs in the alkaline-earth tungstates and in alkaline perrhenates **[48]**. The above similarities between the modes of these compounds are likely due to similar electronegativities of the Ag and Pb cations, that lead to negligible charge-transfer effects between the A$^{2+}$ cation and the BO$_4^{2-}$ anion and consequently to low frequencies and pressure coefficients of the internal BO$_4$ modes due to the weakness of the W-O bond as already discussed in PbWO$_4$.

The main differences between the scheelite phases of the three compounds are: 1) the frequencies of the internal stretching modes of ReO$_4$ tetrahedra are higher than those of the WO$_4$ and MoO$_4$ tetrahedra, as already noted in the discussion of the ionicity of PbWO$_4$. This is due to the larger stretching force constant of ReO$_4$ tetrahedra as a consequence of the shorter Re-O bond distances compared to the W-O and Mo-O bond distances; 2) the pressure coefficients of all external modes are somewhat larger in AgReO$_4$. In fact the lowest T(B$_g$) mode does not exhibit a negative pressure coefficient in AgReO$_4$ as in the studied tungstates and molybdates. This is in agreement with the larger stability of the scheelite structure in the perrhenate; 3) the scheelite $\nu_1$(A$_g$) mode shows a very small pressure coefficient in PbMoO$_4$ and AgReO$_4$, thus suggesting a negligible decrease of the B-O bond distance in these compounds.

We believe that these similarities and differences in the scheelite phases of the three compounds can be also traced in the high-pressure Raman spectra of PbWO$_4$ and AgReO$_4$. Phase transitions between 11 and 14 GPa are reported in AgReO$_4$ that exhibits a high-frequency mode at 942 cm$^{-1}$ whose intensity decrease strongly above 14 GPa similarly to the ferguson ite mode of PbWO$_4$ located around 870 cm$^{-1}$. Furthermore, two modes located at 275 and 301 cm$^{-1}$ at 14 GPa in AgReO$_4$ exhibit a nearly zero-pressure coefficient like the PbWO$_4$–III modes located at 288 and 305 cm$^{-1}$ at 13.7 GPa. Finally, we must note that there is a broad band composed of three peaks near 220 cm$^{-1}$ in AgReO$_4$ that resembles that of the PbWO$_4$-III phase near 200 cm$^{-1}$. A closer comparison cannot be made due to the lack of more detailed spectra of AgReO$_4$, especially in the 450 to 850 cm$^{-1}$ region.

The main differences between the high-pressure phases of the three compounds are the following ones. Assuming that in PbMoO$_4$ there is a phase transition to the ferguson ite phase around 9.5 GPa, and that in AgReO$_4$ there is a first phase transition to the ferguson ite phase around 11 GPa and a second phase transition to the PbWO$_4$-III phase around 14 GPa, a striking difference between the high-pressure Raman spectrum of the three compounds is that the change in frequency of the ferguson ite high-frequency stretching A$_g$ mode with respect to the scheelite $\nu_1$(A$_g$) mode is very small in PbMoO$_4$ and in AgReO$_4$ (at most 4 cm$^{-1}$), unlike in PbWO$_4$. Therefore, the ferguson ite phases of PbMoO$_4$ and AgReO$_4$ resemble more those of alkaline-earth tungstates and likely retain the tetrahedral B cation coordination **[10]**. This result suggests that the jump between these two modes at the scheelite-to-ferguson ite phase transition in PbWO$_4$ due to the internal distortions revealed by *ab initio* calculations is particular for PbWO$_4$ and is not usually found in other scheelites. On the other hand, two new and strong peaks are observed above 14 GPa in the Raman spectrum of $AgReO_4$. One is at 71 $cm^{-1}$ and the other is around 951-953 $cm^{-1}$. These two modes maintain their intensities at the highest pressures, while the 942 $cm^{-1}$ mode assigned to the fergusonite phase fades. We think that these two peaks could correspond to the $PbWO_4$-III phase of $AgReO_4$. However, the observation of only one strong high-frequency peak in this phase, as compared to the two strong high-frequency peaks observed in $PbWO_4$, which could be due to the limited resolution of the Raman spectra of $AgReO_4$, lead us to think about the possibility of a transition to another monoclinic structure with less formula units per unit cell, like the $LaTaO_4$ structure. The hypothesis stated above is valid for $PbMoO_4$, which exhibits three modes near the scheelite $\nu_1(A_g)$ mode and two more modes near the scheelite $\nu_3$ modes after the phase transition at 9.5 GPa. The three modes near the scheelite $\nu_1(A_g)$ mode are compatible with the assignment of one of them to the fergusonite phase, and the remaining two modes to the $PbWO_4$-III phase. In any case, further Raman studies in $PbMoO_4$ and $AgReO_4$ are needed to fully understand the pressure behavior of these two scheelites.

**4. Conclusions**

We performed RT Raman scattering measurements under pressure in $PbWO_4$ up to 17 GPa. The frequency pressure dependence of all the first-order modes of the scheelite phase have been measured. We have observed the onset of the scheelite-to-$PbWO_4$-III phase transition around 6.2 GPa in good agreement with the theoretically calculated value of the I/III coexistence pressure (5.3 GPa) **[12]**. Our measurements show that the transition to the $PbWO_4$-III phase is not completed up to 14.6 GPa. We have also found a scheelite-to-fergusonite phase transition at 7.9 GPa in complete agreement with earlier experimental and theoretical studies **[12]**. In summary, we have observed the following structural sequences in PbWO$_4$: 1) scheelite from 1 atm to 6.2 GPa; 2) scheelite+PbWO$_4$-III (partial transition) between 6.2 and 7.9 GPa; 3) scheelite + PbWO$_4$-III + fergusonite between 7.9 and 9.0 GPa; 4) PbWO$_4$-III + fergusonite from 9.5 GPa to 14.6 GPa; 5) PbWO$_4$-III above 15 GPa. On decreasing pressure from 17 GPa the scheelite structure of PbWO$_4$ is recovered below 5 GPa in agreement with ADXRD measurements **[12]**.

The observation of the scheelite-to-fergusonite transition at pressures above that for the scheelite-to-PbWO$_4$-III transition can only be explained by the kinetic hindrance of the reconstructive scheelite-to-PbWO$_4$-III phase transition due to its slow kinetics likely caused by an activation barrier, and the displacive second-order nature of the scheelite-to-fergusonite phase transition, as already pointed out in **Refs. 10, 12** and **43**. As a consequence of this, we have found the coexistence of the scheelite, fergusonite and PbWO$_4$-III phases in the pressure range of 7.9 to 9 GPa and the coexistence of the last two phases up to 14.6 GPa, despite most of the fergusonite modes disappear above 12.6 GPa. These results allow us to understand previous x-ray diffraction results **[12]**. The Raman peaks of the PbWO$_4$-III phase appear before those of the fergusonite, but once the fergusonite phase appears its Raman peaks are stronger than those of PbWO$_4$-III and dominate the spectrum up to 12.6 GPa, pressure at which the PbWO$_4$-III phase dominates over the fergusonite. This result explains why in the ADXRD study **[12]** the fergusonite phase was observed at 9 GPa and why the ADXRD pattern of the PbWO$_4$-III phase appears only after the fergusonite phase extinguishes above 14.6 GPa.

Finally, we have performed *ab initio* lattice dynamics calculations of PbWO$_4$ at selected pressures in the scheelite, fergusonite, and PbWO$_4$-III phases. Our calculated frequencies in the three structures agree with the frequencies of the observed Raman modes and have allowed the assignment and discussion of the nature of many modes in the three phases. We have found that PbWO$_4$ under pressure behaves in a similar way than BaWO$_4$ **[10]**, and AgReO$_4$ **[48]**. With the aid of Hardcastle and Wachs', and of Brown and Wu's formulae and *ab initio* calculations, we have shown that the WO$_4$ tetrahedra in stolzite become distorted in fergusonite-PbWO$_4$. The distortion leads to a change of interatomic W-O distances in fergusonite-PbWO$_4$ that suggests that this phase acts as a bridge between fourfold and sixfold W coordination. The distortion of the W-O bonds of the fergusonite phase in PbWO$_4$, evidenced by *ab initio* calculations, explains the big decrease of the frequency of one of the Raman stretching modes of this phase, and why the scheelite-to-fergusonite transition is observed in XANES measurements while the fergusonite-to-PbWO$_4$-III transition is faintly observed in XANES measurements **[12]**. Furthermore, we have shown that the PbWO$_4$-III phase has octahedral coordination for the W cation, and that the WO$_6$ octahedra in the fergusonite and PbWO$_4$-III phases can be regarded as almost independent units.

**Acknowledgments**

The authors thank P. Lecoq (CERN) for providing the PbWO$_4$ crystals used in this study. This work was made possible through financial support of the MCYT of Spain under grants No.: MAT2004-05867-C03-01/03, and MAT2002-04539-C02-02. F. J. M. acknowledges financial support by the "*Programa Incentivo a la Investigación de la U.P.V.*". D.E. and N.G. acknowledge the financial support from the MCYT of Spain through the "Ramon y Cajal" program. The use of the computational resources of the Barcelona Supercomputing Center (Mare Nostrum) for the DFT calculations is also gratefully acknowledged. J.L.S., A.M., and P. R-H. acknowledge the financial support from the Consejería de Educación del Gobierno Autónomo de Canarias (PI2003/074).

## Tables.

**Table I.** *Ab initio* calculated and experimental frequencies, pressure coefficients, and Grüneisen parameters of the Raman modes of scheelite-PbWO$_4$ at 1 atm. For obtaining the Grüneisen parameter, $\gamma = B_0/\omega(0) \cdot d\omega/dP$, we have taken the bulk modulus of scheelite-PbWO$_4$, $B_0$ = 66 GPa **[12]**.

| Peak/ mode | $\omega(0)$ cm$^{-1}$ | $d\omega/dP$ cm$^{-1}$/GPa | $\gamma$ | $\omega(0)^c$ cm$^{-1}$ | $d\omega/dP^c$ cm$^{-1}$/GPa | $\gamma^c$ |
|---|---|---|---|---|---|---|
| T(B$_g$) | 58 | -1.1 | -1.30$^a$, -1.10$^b$ | 52 | 0.7 | 0.81 |
| T(E$_g$) | 65 | 1.8 | 1.90$^a$, 3.20$^b$ | 64 | 2.3 | 2.37 |
| T(B$_g$) | 77 | 3.3 | 2.80$^a$, 2.20$^b$ | 79 | 4.4 | 3.67 |
| T(E$_g$) | 90 | 2.3 | 1.60$^a$, 1.70$^b$ | 92 | 4.6 | 3.30 |
| R(A$_g$) | 178 | 3.3 | 1.20$^a$, 1.40$^b$ | 191 | 3.3 | 1.14 |
| R(E$_g$) | 193 | 4.2 | 1.40$^a$ | 193 | 4.6 | 1.57 |
| $\nu_2$(A$_g$) | 323 | 1.9 | 0.40$^a$ | 310 | 2.1 | 0.45 |
| $\nu_2$(B$_g$) | 328 | 2.1 | 0.40$^a$, 0.60$^b$ | 311 | 3.0 | 0.64 |
| $\nu_4$(B$_g$) | 357 | 2.8 | 0.50$^a$, 0.60$^b$ | 350 | 2.7 | 0.51 |
| $\nu_4$(E$_g$) | 362 | 2.7 | 0.50$^a$ | 351 | 2.8 | 0.53 |
| $\nu_3$(E$_g$) | 752 | 2.4 | 0.20$^a$, 0.30$^b$ | 750 | 3.1 | 0.27 |
| $\nu_3$(B$_g$) | 766 | 0.9 | 0.08$^a$, 0.05$^b$ | 758 | 1.7 | 0.15 |
| $\nu_1$(A$_g$) | 906 | 0.8 | 0.06$^a$, 0.08$^b$ | 890 | 1.4 | 0.10 |

$^a$ This work, $^b$**Ref. 9**, $^c$*Ab initio* calculations.

**Table II.** Frequencies, pressure coefficients, and Grüneisen parameters of the calculated IR modes in scheelite-PbWO$_4$ at 1 atm. Experimentally measured IR-active modes at RT are also reported for comparison. The values for the silent B$_u$ modes are also shown for completeness.

| Peak/mode | ω(0) (cm$^{-1}$) | dω/dP (cm$^{-1}$/GPa) | γ | ω(0) exp. (cm$^{-1}$) |
|---|---|---|---|---|
| T(A$_u$) | 0 | - | - | 0 |
| T(E$_u$) | 0 | - | - | 0 |
| T(E$_u$) | 54 | 7.0 | 8.50 | 58[a] |
| T(A$_u$) | 71 | 7.1 | 6.60 | 73[a] |
| R(E$_u$) | 122 | -0.8 | -0.43 | 104[a] |
| R(B$_u$) | 222 | 0.3 | 0.09 | |
| ν$_4$(A$_u$) | 243 | -2.1 | -0.57 | 251[a] |
| ν$_4$(E$_u$) | 275 | 1.8 | 0.43 | 288[a] |
| ν$_2$(A$_u$) | 370 | 2.5 | 0.45 | 384[a] |
| ν$_2$(B$_u$) | 385 | 5.2 | 0.89 | |
| ν$_3$(A$_u$) | 746 | 2.4 | 0.21 | 756[a], 757[c] |
| ν$_3$(E$_u$) | 754 | 2.6 | 0.23 | 764[a], 771[c] |
| ν$_1$(B$_u$) | 891 | 2.0 | 0.15 | 862[b], 850[c] |

[a] Ref. 25, [b] Ref. 32, [c] Ref. 26

**Table III.** Frequencies, pressure coefficients, and relative pressure coefficients of the Raman modes in fergusonite-PbWO$_4$ at 9 GPa. The fergusonite frequencies and pressure coefficients obtained after *ab initio* calculations at 9 GPa are also given for comparison.

| Peak/mode | ω(9) cm$^{-1}$ | dω/dP cm$^{-1}$/GPa | 1/ω·dω/dP | ω(9)$^a$ cm$^{-1}$ | dω/dP$^a$ cm$^{-1}$/GPa |
|---|---|---|---|---|---|
| F1 (B$_g$) | 44(1) | 0.8(1) | 0.018 | 65 | 1.9 |
| F2 (A$_g$) | 53(1) | 0.4(2) | 0.008 | 78 | 1.5 |
| F3 (B$_g$) | 85(1) | 1.4(3) | 0.016 | 84 | 1.6 |
| F4 (B$_g$) | 136(1) | 3.7(3) | 0.027 | 134 | 4.2 |
| F5 (A$_g$) | 144(1) | 5.5(5) | 0.038 | 150 | 5.0 |
| F6 (B$_g$) | 158(1) | 4.0(1) | 0.025 | 173 | 3.6 |
| F7 (B$_g$) | | | | 206 | 7.5 |
| F8 (B$_g$) | | | | 247 | 7.0 |
| F9 (A$_g$) | 261(1) | 1.9(1) | 0.007 | 295 | 3.7 |
| F10 ν$_2$(A$_g$) | 320(1) | 1.9(3) | 0.006 | 303 | 4.0 |
| F11 ν$_2$(A$_g$) | 352(1) | 1.7(2) | 0.005 | 365 | 3.2 |
| F12 ν$_4$(A$_g$) | 396(1) | 0.9(1) | 0.002 | 377 | 2.5 |
| F13 ν$_4$(B$_g$) | 446(1) | 3.2(2) | 0.007 | 440 | 5.4 |
| F14 ν$_4$(B$_g$) | 471(1) | 5.6(5) | 0.012 | 467 | 7.9 |
| F15 ν$_3$(A$_g$) | 693(1) | 1.0(2) | 0.001 | 661 | 6.8 |
| F16 ν$_3$(B$_g$) | 725(2) | 3.3(4) | 0.005 | 689 | -0.8 |
| F17 ν$_3$(B$_g$) | 779(1) | 3.6(5) | 0.005 | 752 | 4.4 |
| F18 ν$_1$(A$_g$) | 872(1) | 0.9(1) | 0.001 | 846 | 1.0 |

$^a$*Ab initio* calculations.

**Table IV.** IR mode frequencies and pressure coefficients in fergusonite-PbWO$_4$ obtained from *ab initio* calculations at 9 GPa.

| Peak/mode | ω(9) cm$^{-1}$ | dω/dP cm$^{-1}$/GPa |
|---|---|---|
| F1(A$_u$) | 0 | - |
| F2(B$_u$) | 0 | - |
| F3(B$_u$) | 0 | - |
| F4(A$_u$) | 85 | 2.9 |
| F5(B$_u$) | 91 | 1.2 |
| F6(B$_u$) | 114 | 3.4 |
| F7(A$_u$) | 161 | 6.8 |
| F8(B$_u$) | 210 | 6.9 |
| F9(B$_u$) | 215 | 1.3 |
| F10(A$_u$) | 295 | 3.1 |
| F11(A$_u$) | 306 | 7.8 |
| F12(A$_u$) | 356 | 9.1 |
| F13(B$_u$) | 388 | 2.7 |
| F14(B$_u$) | 463 | 7.6 |
| F15(B$_u$) | 610 | 2.9 |
| F16(B$_u$) | 705 | 2.2 |
| F17(A$_u$) | 720 | 3.5 |
| F18(A$_u$) | 854 | 6.1 |

**Table V.** Frequencies and pressure coefficients of the Raman modes observed in the PbWO$_4$-III phase at 13.7 GPa.

| Peak /mode | ω(13.7) cm$^{-1}$ | dω/dP cm$^{-1}$/GPa | Peak /mode | ω(13.7) cm$^{-1}$ | dω/dP cm$^{-1}$/GPa |
|---|---|---|---|---|---|
| P1 | 46(1) | - | P20 | 400(1) | - |
| P2 | 61(1) | 0.2(2) | P21 | 417(2) | 2.1(3) |
| P3 | 68(1) | 0.0(2) | P22 | 430(1) | 2.2(2) |
| P4 | 69(1) | 1.2(2) | P23 | 490(3) | 1.1(3) |
| P5 | 73(1) | 0.7(1) | P24 | 501(2) | 1.1(2) |
| P6 | 84(1) | 0.0(2) | P25 | 542(2) | 4.6(5) |
| P7 | 120(1) | 0.6(1) | P26 | 564(3) | 5.4(6) |
| P8 | 123(1) | 0.0(2) | P27 | 605(3) | 1.6(2) |
| P9 | 130(2) | 0.6(1) | P28 | 672(3) | 0.2(1) |
| P10 | 145(2) | 4.5(9) | P29 | 693(2) | 0.7(1) |
| P11 | 189(1) | 0.9(2) | P30 | 710(2) | 1.0(2) |
| P12 | 201(1) | 1.2(2) | P31 | 773(2) | 1.6(2) |
| P13 | 208(1) | 1.3(2) | P32 | 799(2) | 1.4(2) |
| P14 | 222(5) | 0.2(2) | P33 | 825(3) | 1.9(3) |
| P15 | 246(1) | 1.3(3) | P34 | 899(1) | 1.5(3) |
| P16 | 266(1) | -2.2(5) | P35 | 914(1) | 2.5(3) |
| P17 | 288(1) | 0.0(3) | P36 | 935(2) | 0.8(1) |
| P18 | 305(1) | 0.0(3) | P37 | 949(2) | 1.5(4) |
| P19 | 390(1) | 0.0(3) | | | |

**Table VI.** Raman mode symmetries and frequencies in the PbWO$_4$-III phase as obtained from *ab initio* calculations at 8.7 GPa.

| Mode (sym) | ω(8.7) (cm$^{-1}$) | Mode (sym) | ω(8.7) (cm$^{-1}$) | Mode (sym) | ω(8.7) (cm$^{-1}$) | Mode (sym) | ω(8.7) (cm$^{-1}$) |
|---|---|---|---|---|---|---|---|
| R1(B$_g$) | 43 | R19(A$_g$) | 130 | R37(A$_g$) | 269 | R55(A$_g$) | 495 |
| R2(A$_g$) | 44 | R20(B$_g$) | 136 | R38(B$_g$) | 283 | R56(B$_g$) | 503 |
| R3(B$_g$) | 52 | R21(B$_g$) | 142 | R39(B$_g$) | 306 | R57(A$_g$) | 575 |
| R4(A$_g$) | 56 | R22(A$_g$) | 153 | R40(A$_g$) | 309 | R58(B$_g$) | 576 |
| R5(A$_g$) | 59 | R23(B$_g$) | 170 | R41(A$_g$) | 321 | R59(A$_g$) | 644 |
| R6(A$_g$) | 63 | R24(A$_g$) | 171 | R42(B$_g$) | 332 | R60(B$_g$) | 645 |
| R7(B$_g$) | 66 | R25(A$_g$) | 173 | R43(A$_g$) | 340 | R61(A$_g$) | 694 |
| R8(B$_g$) | 69 | R26(B$_g$) | 179 | R44(B$_g$) | 341 | R62(B$_g$) | 697 |
| R9(A$_g$) | 72 | R27(A$_g$) | 191 | R45(B$_g$) | 362 | R63(B$_g$) | 718 |
| R10(A$_g$) | 74 | R28(B$_g$) | 196 | R46(A$_g$) | 368 | R64(A$_g$) | 719 |
| R11(B$_g$) | 81 | R29(B$_g$) | 213 | R47(A$_g$) | 380 | R65(A$_g$) | 730 |
| R12(A$_g$) | 87 | R30(A$_g$) | 215 | R48(B$_g$) | 390 | R66(B$_g$) | 761 |
| R13(B$_g$) | 91 | R31(B$_g$) | 232 | R49(A$_g$) | 391 | R67(A$_g$) | 774 |
| R14(B$_g$) | 101 | R32(B$_g$) | 237 | R50(B$_g$) | 400 | R68(B$_g$) | 802 |
| R15(A$_g$) | 104 | R33(A$_g$) | 244 | R51(B$_g$) | 436 | R69(A$_g$) | 857 |
| R16(B$_g$) | 109 | R34(A$_g$) | 255 | R52(A$_g$) | 438 | R70(B$_g$) | 869 |
| R17(A$_g$) | 118 | R35(B$_g$) | 258 | R53(A$_g$) | 467 | R71(A$_g$) | 912 |
| R18(B$_g$) | 121 | R36(A$_g$) | 259 | R54(B$_g$) | 472 | R72(B$_g$) | 919 |

**Table VII.** IR mode symmetries and frequencies in the PbWO$_4$-III phase as obtained from *ab initio* calculations at 8.7 GPa.

| Mode (sym) | ω(8.7) (cm$^{-1}$) | Mode (sym) | ω(8.7) (cm$^{-1}$) | Mode (sym) | ω(8.7) (cm$^{-1}$) | Mode (sym) | ω(8.7) (cm$^{-1}$) |
|---|---|---|---|---|---|---|---|
| I1(A$_u$) | 0 | I19(A$_u$) | 105 | I37(B$_u$) | 283 | I55(A$_u$) | 442 |
| I2(B$_u$) | 0 | I20(A$_u$) | 118 | I38(A$_u$) | 284 | I56(B$_u$) | 447 |
| I3(B$_u$) | 0 | I21(A$_u$) | 135 | I39(B$_u$) | 296 | I57(A$_u$) | 543 |
| I4(B$_u$) | 36 | I22(B$_u$) | 140 | I40(A$_u$) | 302 | I58(B$_u$) | 548 |
| I5(A$_u$) | 38 | I23(B$_u$) | 147 | I41(A$_u$) | 317 | I59(B$_u$) | 605 |
| I6(A$_u$) | 47 | I24(A$_u$) | 148 | I42(B$_u$) | 322 | I60(A$_u$) | 606 |
| I7(B$_u$) | 61 | I25(A$_u$) | 177 | I43(B$_u$) | 328 | I61(A$_u$) | 705 |
| I8(B$_u$) | 61 | I26(B$_u$) | 178 | I44(A$_u$) | 338 | I62(B$_u$) | 706 |
| I9(A$_u$) | 62 | I27(A$_u$) | 183 | I45(A$_u$) | 359 | I63(A$_u$) | 720 |
| I10(B$_u$) | 69 | I28(B$_u$) | 185 | I46(B$_u$) | 361 | I64(B$_u$) | 727 |
| I11(A$_u$) | 70 | I29(B$_u$) | 202 | I47(B$_u$) | 374 | I65(B$_u$) | 742 |
| I12(A$_u$) | 72 | I30(A$_u$) | 213 | I48(A$_u$) | 380 | I66(A$_u$) | 778 |
| I13(B$_u$) | 77 | I31(A$_u$) | 220 | I49(B$_u$) | 388 | I67(A$_u$) | 779 |
| I14(B$_u$) | 84 | I32(B$_u$) | 228 | I50(A$_u$) | 392 | I68(B$_u$) | 804 |
| I15(A$_u$) | 92 | I33(A$_u$) | 245 | I51(A$_u$) | 397 | I69(B$_u$) | 862 |
| I16(B$_u$) | 94 | I34(B$_u$) | 248 | I52(B$_u$) | 400 | I70(A$_u$) | 864 |
| I17(A$_u$) | 98 | I35(B$_u$) | 259 | I53(B$_u$) | 418 | I71(B$_u$) | 898 |
| I18(B$_u$) | 104 | I36(A$_u$) | 273 | I54(A$_u$) | 438 | I72(A$_u$) | 911 |

**Figure captions**

**Fig. 1.** RT Raman spectra of scheelite-PbWO$_4$ at different pressures between 1 atm and 8 GPa. The dashed line indicates the position of a plasma line of Ar$^+$ at 104 cm$^{-1}$ used for calibration of Raman spectra. Arrows indicate the position of the silicon oil modes at different pressures. Inset shows a detail of the Raman spectrum at 1 atm in the region of the $\nu_4(B_g)$ and $\nu_4(E_g)$ modes. Exclamation marks indicate the Raman peaks assigned to the PbWO$_4$-III phase. Asterisks indicate the Raman peaks assigned to the fergusonite phase. The *ab initio* calculated frequencies of the scheelite Raman modes at 1 atm are marked at the bottom.

**Fig. 2.** Detail of the Raman spectra of scheelite-PbWO$_4$ at different pressures around the $\nu_2(B_g)$ mode. Raman spectra above 0.3 GPa have been shifted by -1.97, -5.53, -8.45, -11.21, -13.80, and -15.42 cm$^{-1}$, respectively, in order to bring the mode at 328 cm$^{-1}$ (long dashed line) into coincidence. The short dashed line indicates the relative evolution of the low-frequency mode at 323 cm$^{-1}$ at 1 atm with respect to the 328 cm$^{-1}$ line. The arrow indicates a mode of the fergusonite phase which begins to appear in the spectrum around 8 GPa.

**Fig. 3.** Pressure dependence of the Raman mode frequencies of the scheelite (solid circles), fergusonite (empty squares) and PbWO$_4$-III (solid triangles) phases of PbWO$_4$ up to 17 GPa. Empty circles show the pressure behaviour of the two silicon oil Raman modes observed. Dotted lines show the onset of the scheelite-to-PbWO$_4$-III phase transition and of the scheelite-to-fergusonite transition. Dashed line indicates the pressure for the completion of the transition to the PbWO$_4$-III phase. The solid lines are guides to the eye.

**Fig. 4.** RT Raman spectra of PbWO$_4$ at different pressures between 6 and 17 GPa. Exclamation marks indicate the Raman peaks assigned to the PbWO$_4$-III phase. Asterisks indicate the Raman peaks assigned to the fergusonite phase. Arrows indicate the position of the silicon oil modes at different pressures. The dashed line indicates the position of a plasma line of Ar$^+$ at 104 cm$^{-1}$ used for calibration of Raman spectra.

**Fig. 5.** Detail of the Raman spectra of the fergusonite and PbWO$_4$-III phases of PbWO$_4$ at different pressures at low frequencies. The dashed line indicates the position of a plasma line of Ar$^+$ at 104 cm$^{-1}$ used for calibration of Raman spectra. Arrows indicate the position of scheelite modes. Exclamation marks indicate the Raman peaks assigned to the PbWO$_4$-III phase. Asterisks indicate the Raman peaks assigned to the fergusonite phase.

**Fig. 6.** Detail of the Raman spectra of the fergusonite and PbWO$_4$-III phases of PbWO$_4$ at different pressures at medium frequencies. Arrows indicate the position of scheelite modes. Exclamation marks indicate the Raman peaks assigned to the PbWO$_4$-III phase. Asterisks indicate the Raman peaks assigned to the fergusonite phase.

**Fig. 7.** Detail of the Raman spectra of the fergusonite and PbWO$_4$-III phases of PbWO$_4$ at different pressures at high frequencies. Arrows indicate the position of scheelite modes. Asterisks indicate the position of fergusonite modes. Exclamation marks indicate the Raman peaks assigned to the PbWO$_4$-III phase. Asterisks indicate the Raman peaks assigned to the fergusonite phase. The dashed line indicates the position of one of the silicon oil modes observed at high pressures.

**Fig. 8.** W-O (squares) and Pb-O (circles) distances in the scheelite (solid) and fergusonite (empty) phases of PbWO$_4$ at different pressures, as obtained from *ab initio* total-energy calculations after **Ref. 12**. The dashed line shows the theoretically predicted phase transition pressure between the scheelite and fergusonite phases. The dotted line shows the division of the fergusonite phase with tetrahedral and octahedral coordination for the W ion.

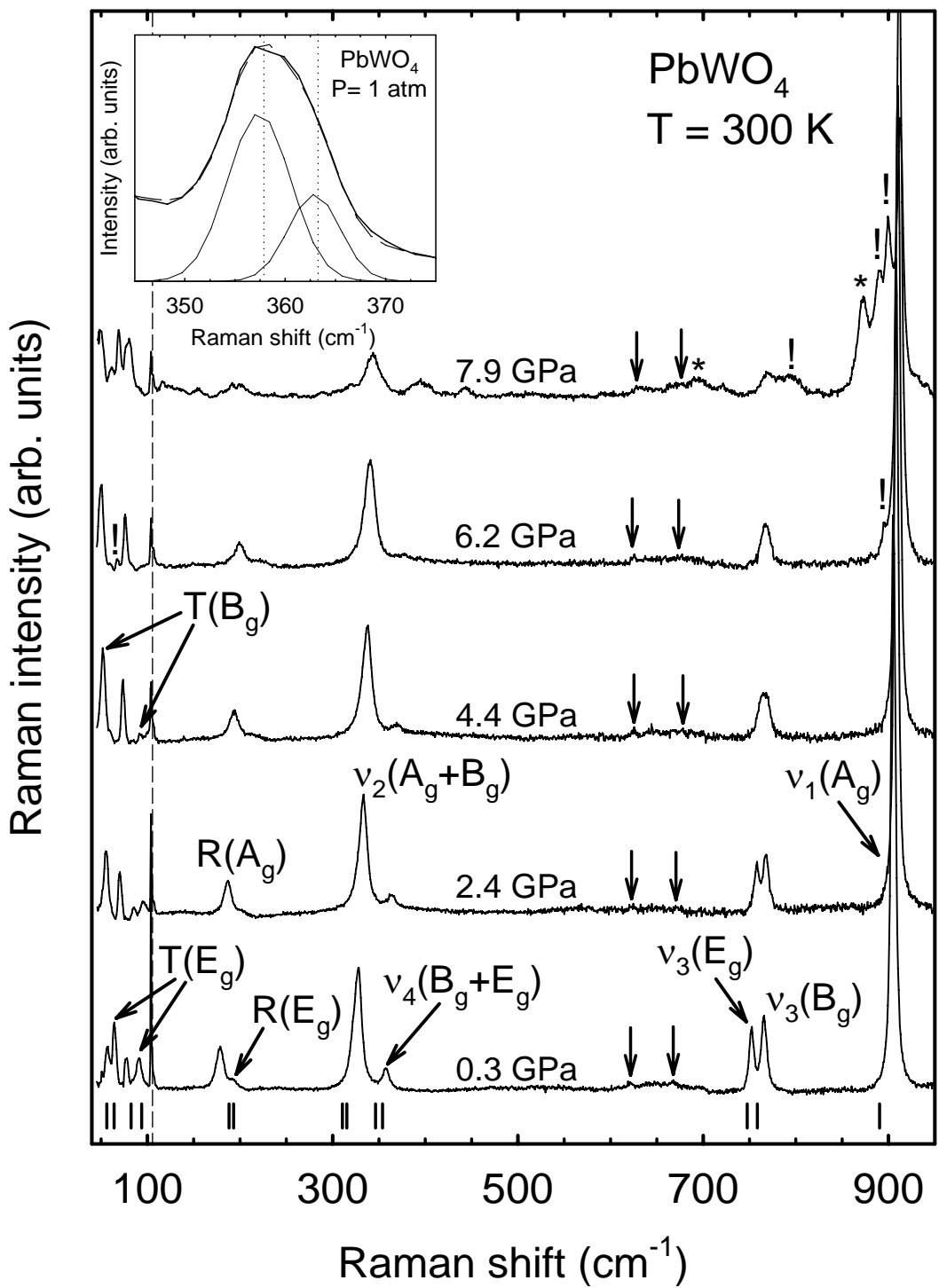

**Figure 1.** F.J. Manjón et al.

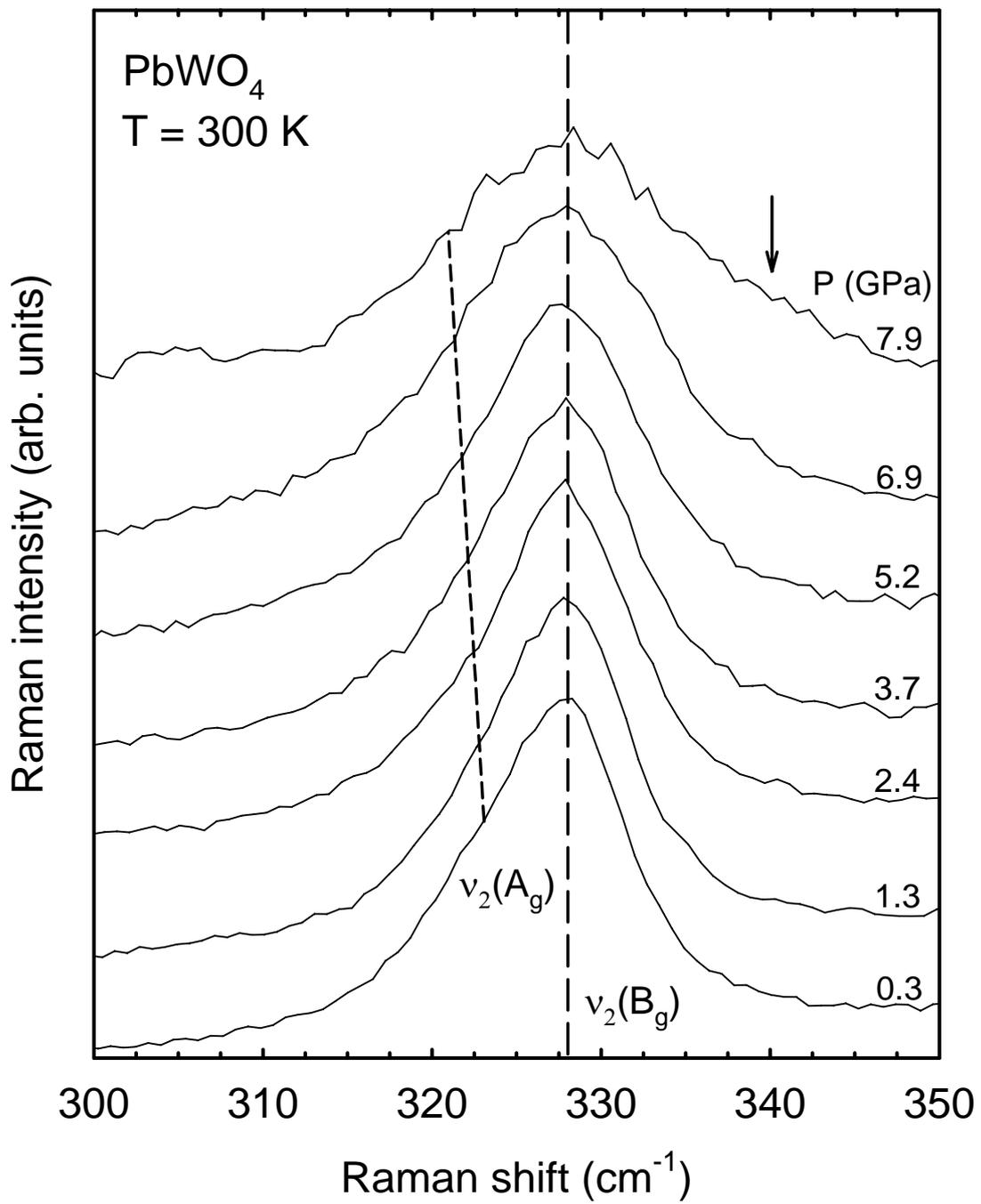

**Figure 2.** F.J. Manjón et al.

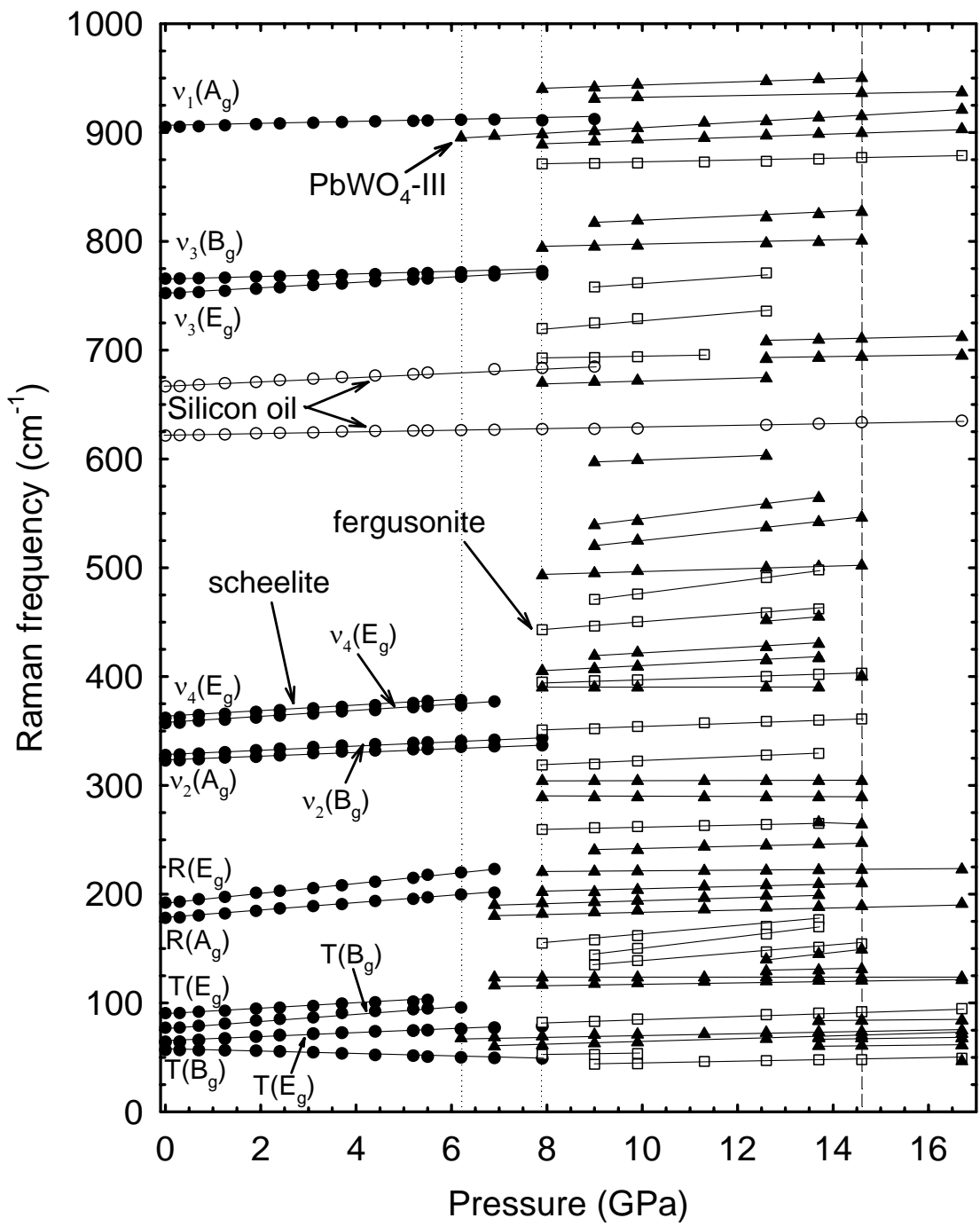

**Figure 3.** F.J. Manjón et al.

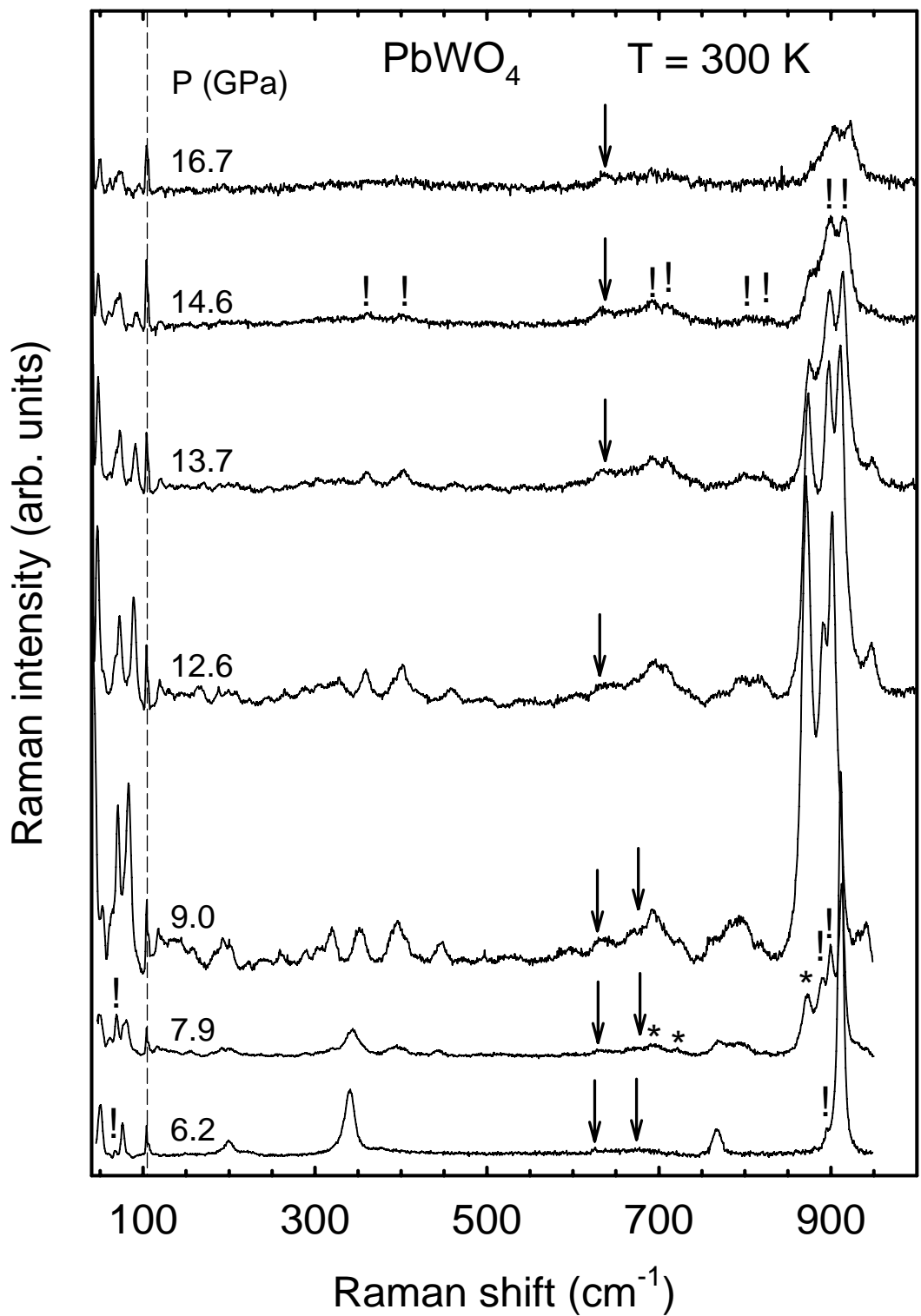

**Figure 4.** F.J. Manjón et al.

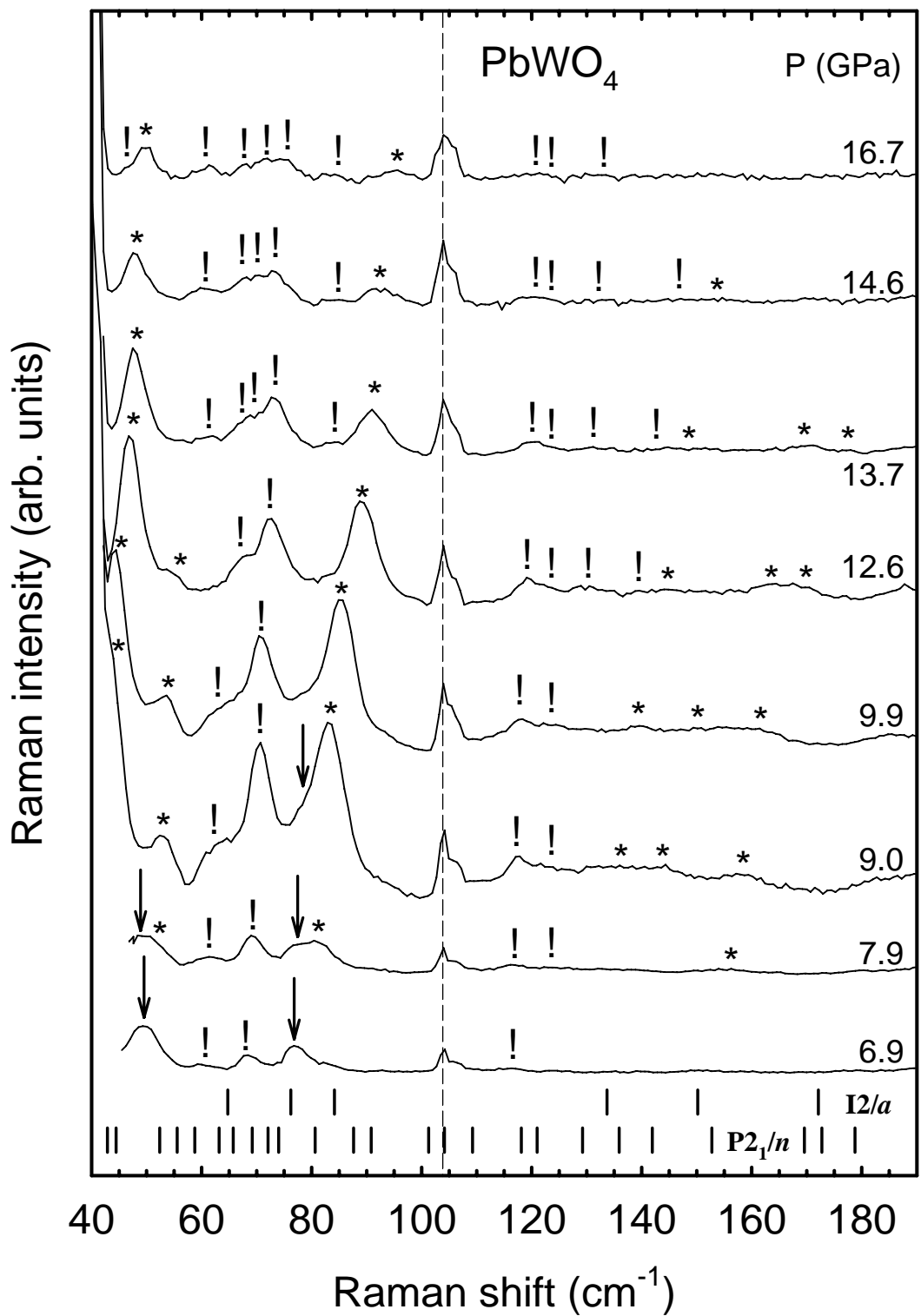

**Figure 5.** F.J. Manjón et al.

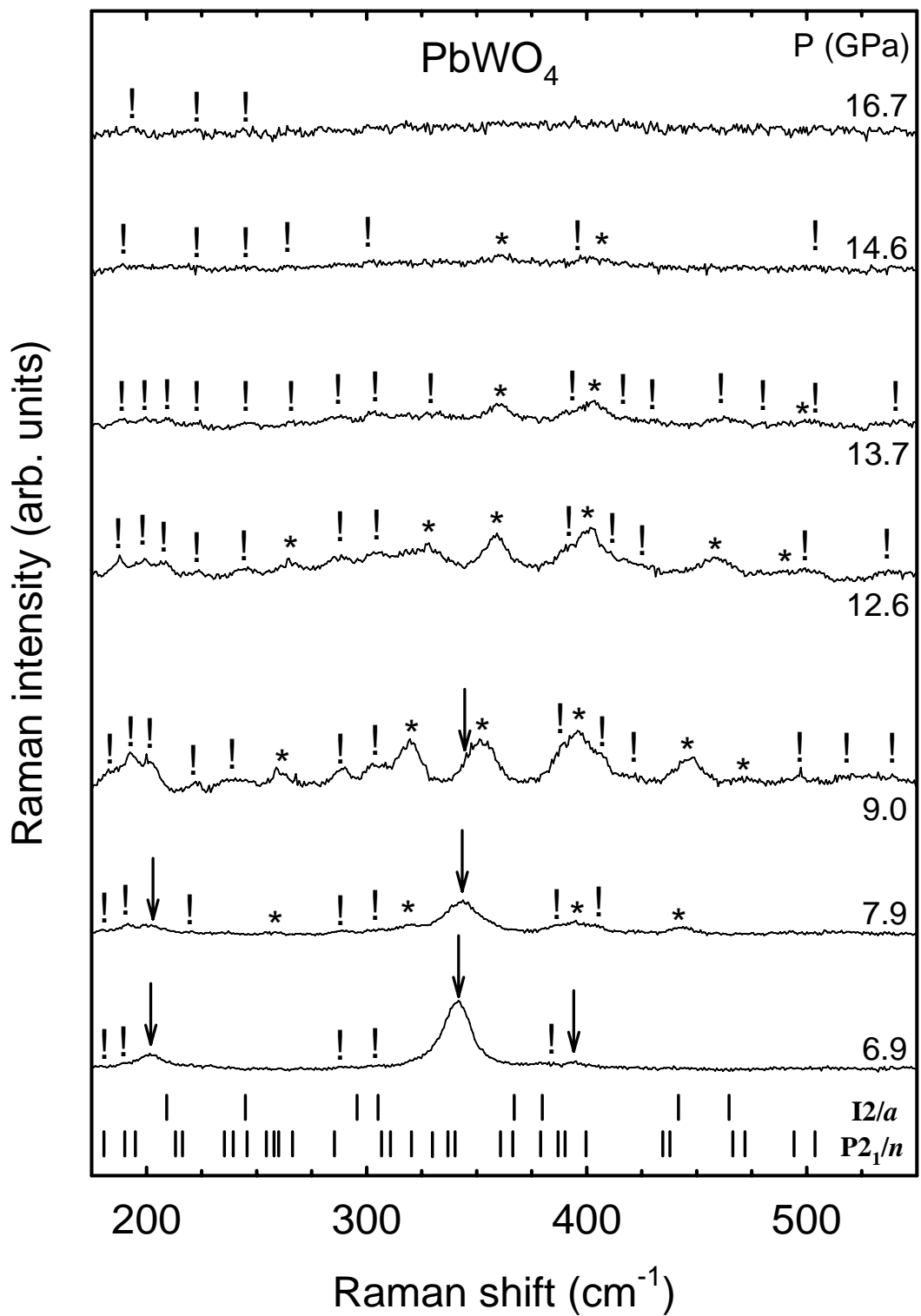

**Figure 6.** F.J. Manjón et al.

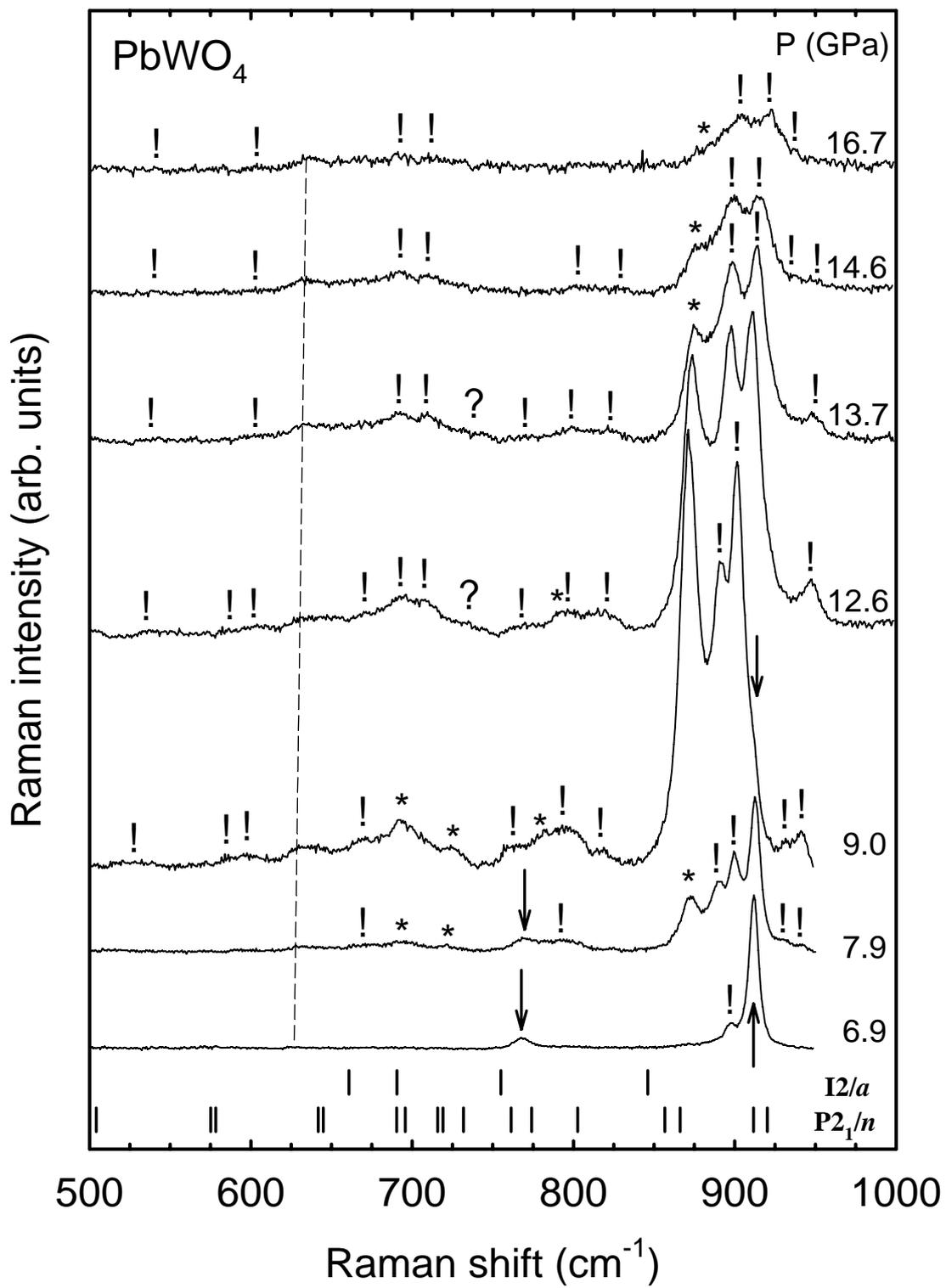

**Figure 7.** F.J. Manjón et al.

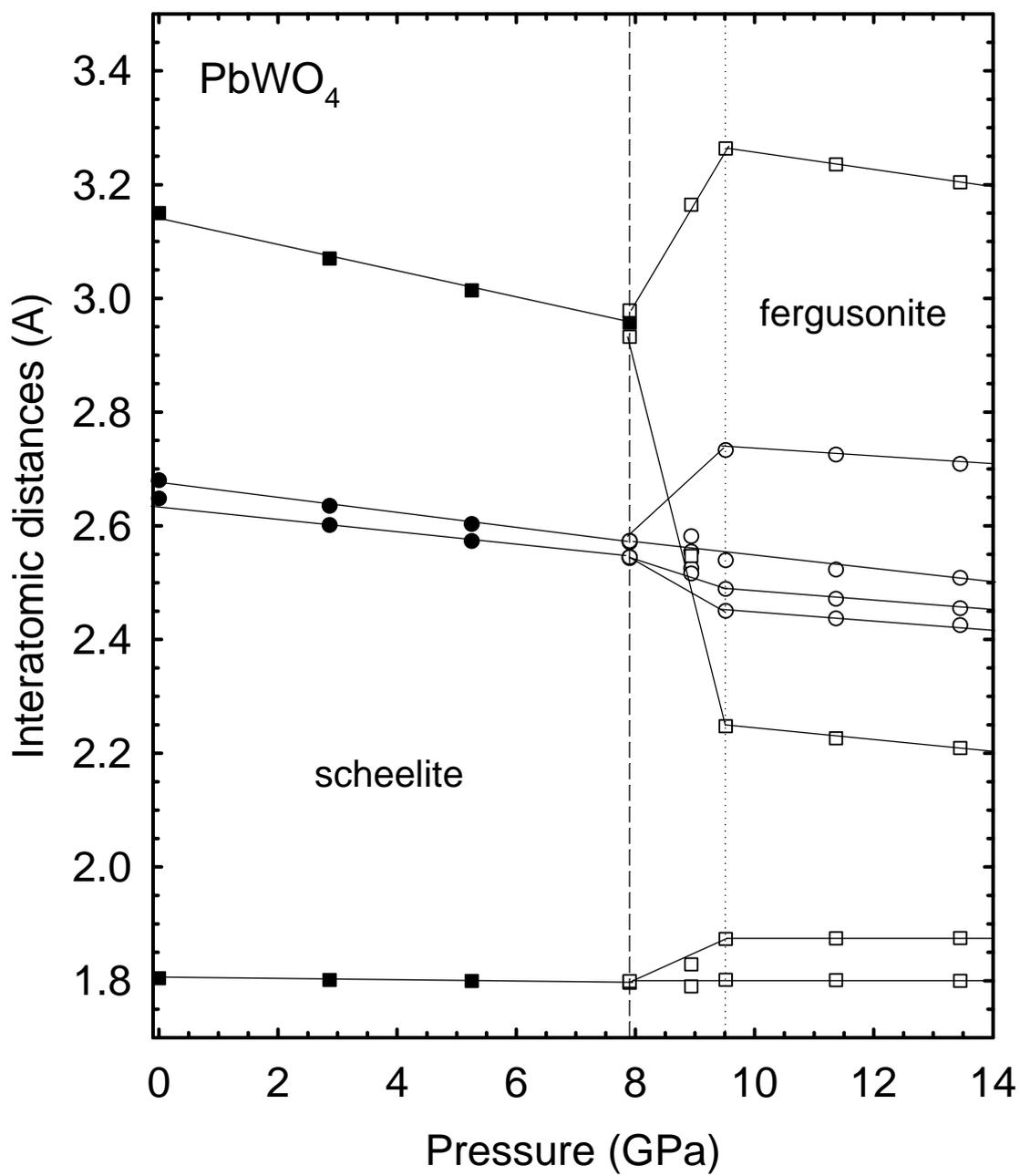

**Figure 8.** F.J. Manjón et al.